\documentclass[reprint,superscriptaddress,amsmath,amssymb,aps,prx,longbibliography]{revtex4-2}

\usepackage{subcaption}
\usepackage{ragged2e} 
\DeclareCaptionJustification{justified}{\justifying}
\usepackage{graphicx}
\graphicspath{{figures/}}

\usepackage{dcolumn}
\usepackage{bm}
\usepackage{braket}

\usepackage[dvipsnames, table]{xcolor}

\usepackage{orcidlink}

\usepackage{hyperref}
\def\colorLinks{black}
\def\colorUrl{blue}
\def\colorCitations{blue}
\hypersetup{
  colorlinks=true,
  linkcolor=\colorLinks,
  urlcolor=\colorUrl,
  citecolor=\colorCitations
}

\usepackage{hyperxmp}

\usepackage{algorithm2e}
\RestyleAlgo{ruled}
\usepackage{algpseudocode}
\SetKwComment{Comment}{// }{}

\begin{document}

\preprint{APS/123-QED}

\title{\texorpdfstring{Dynamical cluster-based strategy for improving tensor network \\ algorithms in quantum circuit simulations}{}}

\author{Andrea De Girolamo\orcidlink{0009-0002-4529-0139}}\email{andrea.degirolamo@phd.unipd.it}
\affiliation{Dipartimento Interateneo di Fisica, Università degli Studi di Bari \& Politecnico di Bari, I-70126 Bari, Italy}
\affiliation{Technical University of Munich, School of \mbox{Computation, Information and Technology, D-85748, Garching, Germany}}
\affiliation{Dipartimento di Fisica e Astronomia \mbox{``Galileo Galilei'', Università degli Studi di Padova, I-35131, Padova, Italy}}
\affiliation{Istituto Nazionale di Fisica Nucleare, Sezione di Padova, I-35131, Padova, Italy}
\author{Paolo Facchi\orcidlink{0000-0001-9152-6515}}
\affiliation{Dipartimento Interateneo di Fisica, Università degli Studi di Bari \& Politecnico di Bari, I-70126 Bari, Italy}
\affiliation{Istituto Nazionale di Fisica Nucleare, Sezione di Bari, I-70126 Bari, Italy}
\author{Peter Rabl\orcidlink{0000-0002-2560-8835}}
\affiliation{Technical University of Munich, TUM School of Natural Sciences, D-85748 Garching, Germany}
\affiliation{Walther-Meißner-Institut, Bayerische Akademie der Wissenschaften, D-85748 Garching, Germany}
\affiliation{Munich Center for Quantum Science and Technology (MCQST), D-80799 Munich, Germany}
\author{\\Saverio Pascazio\orcidlink{0000-0002-7214-5685}}
\affiliation{Dipartimento Interateneo di Fisica, Università degli Studi di Bari \& Politecnico di Bari, I-70126 Bari, Italy}
\affiliation{Istituto Nazionale di Fisica Nucleare, Sezione di Bari, I-70126 Bari, Italy}
\author{Cosmo Lupo\orcidlink{0000-0002-5227-4009}}
\affiliation{Dipartimento Interateneo di Fisica, Università degli Studi di Bari \& Politecnico di Bari, I-70126 Bari, Italy}
\affiliation{Istituto Nazionale di Fisica Nucleare, Sezione di Bari, I-70126 Bari, Italy}
\author{Giuseppe Magnifico\orcidlink{0000-0002-7280-445X}}
\affiliation{Dipartimento Interateneo di Fisica, Università degli Studi di Bari \& Politecnico di Bari, I-70126 Bari, Italy}
\affiliation{Istituto Nazionale di Fisica Nucleare, Sezione di Bari, I-70126 Bari, Italy}

\date{\today}

\begin{abstract}
We optimize matrix-product state-based algorithms for simulating quantum circuits with finite fidelity, specifically the time-evolving block decimation~(TEBD) and the density-matrix renormalization group~(DMRG) algorithms, by exploiting the irregular arrangement of entangling operations in circuits. We introduce a variation of the standard TEBD algorithm, we termed ``cluster-TEBD'', which dynamically arranges qubits into entanglement clusters, enabling the exact contraction of multiple circuit layers in a single time step. Moreover, we enhance the DMRG algorithm by introducing an adaptive protocol, which analyzes the entanglement distribution within each circuit section to be contracted, dynamically adjusting the qubit grouping at each iteration. We analyze the performances of these enhanced algorithms in simulating both stabilizer and nonstabilizer random-structured quantum circuits, with up to $1000$ qubits and $100$ layers of Clifford and non-Clifford gates, and in simulating Shor's quantum algorithm with up to hundreds of thousands of layers. Our findings show that, even with reasonable computational resources per task, cluster-based approaches can significantly speed up simulations of large-sized quantum circuits and improve the fidelity of the final states. 

\end{abstract}

\maketitle

\section{Introduction}
\label{sec:Intro}
Quantum computing promises significant performance advantages over classical algorithms, particularly for tasks such as search~\cite{Grover1996} and factorization~\cite{Shor1999}. However, quantum computers still face serious challenges, as quantum states are vulnerable to physical effects, such as noise and decoherence, which lead to loss of information over time~\cite{NielsenChuang2010,Preskill2018NISQ}. Although significant progress has been made recently in designing efficient error-correction codes~\cite{Google2024QECbelowSCthreshold}, accurately executing long-time evolutions on quantum computers remains a significant challenge. 

At the same time, simulating a quantum system on a classical computer poses its own challenges. In fact, a generic quantum state of a qubit system contains an amount of information that grows exponentially with the number of qubits. Consequently, the memory required to exactly represent such quantum states quickly exceeds the capabilities of any existing classical computer. This problem can be mitigated through the use of tensor network techniques~\cite{Orus2014IntroTN,Montangero2018TNReview,RevModPhys.93.045003}. In one-dimensional systems, quantum states can be efficiently represented as matrix product states (MPSs)~\cite{Fannes1992FirstMPS,Perez-Garcia2007MPS}, which require significantly less memory than a full quantum state description, albeit at the cost of introducing approximation errors. By leveraging efficient MPS-based algorithms, such as time-evolving block decimation~(TEBD)~\cite{Vidal2003TEBD} and the density matrix renormalization group~(DMRG)~\cite{White1992DMRG,Schollwock2005DMRGRev}, it is possible to achieve very low approximation errors and numerically address key problems in the study of static and dynamical properties of quantum many-body systems. 

The utility of tensor networks in quantum many-body physics has rapidly extended to quantum computing. While simulating ideal large-scale quantum computers on classical machines is exceedingly challenging, noisy intermediate-scale quantum (NISQ) computers exhibit imperfections that lead to computational errors comparable with the approximation errors of tensor network algorithms~\cite{Zhou2020WhatLimits}. Following Google's claim of achieving quantum supremacy~\cite{Arute2019Sycamore} by simulating a random quantum circuit~\cite{Aaronson2017QSComplexity,Harrow2017QSupReq,Bouland2019RCS, Arute2019Sycamore} with a superconducting quantum device~\cite{Makhlin1999JJqubit, Krantz2019SuperconductingReview}, a series of studies demonstrated that it was possible to simulate the same task with similar or better fidelity on large classical computers with optimized tensor network contraction techniques~\cite{Gray2021HyperExact, Pan2021simulatingsycamore, Pan2022samplingsycamore,Liu2021_304s}. Another work by the team of IBM Quantum showed evidence for the utility of NISQ devices~\cite{Preskill2018NISQ} by simulating the two-dimensional (2D) transverse-field Ising model with a Trotterized quantum circuit on a 127-qubit quantum computer~\cite{Kim2023IBMEagle}. Shortly thereafter, tensor network simulations of the same Hamiltonian were performed with good accuracy on relatively low computational resources, increasing the simulated amount of qubits up to 1121~\cite{Tindall2024IBMTN1, Patra2024IBMTN2}.
 
Optimized classical simulations of many-body dynamics and quantum computing can be seen as complementary in the study of collective quantum effects. Indeed, optimizing numerical methods to simulate large and complex quantum systems could lead to understanding quantum phenomena at large scales, thereby aiding in the development of quantum technologies, such as creating digital twins of experimental quantum platforms~\cite{Jaschke_2024, bidzhiev2023cloudondemandemulationquantum}, or optimizing TEBD and DMRG algorithms to simulate both random quantum circuits and quantum algorithms~\cite{Zhou2020WhatLimits, Ayral2023DMRG_QC, Niedermeier2024QalgsTN}.

In this work, we propose an approach based on entanglement clustering~\cite{Fatima2021GateCluster, Chen2021CircuitPartitioner, cicero2024simulationquantumcomputersreview} to improve tensor network algorithms for simulating quantum circuits with finite fidelity. The main idea is that, by dynamically clustering qubits, according to the pattern of entangling gates in a quantum circuit, we can optimize the simulation of deep quantum circuits with irregular entanglement distribution. 

In particular, we focus on the TEBD and DMRG algorithms. For the former, instead of immediately applying individual gates to the initial MPS, we perform an analysis of the structure of the simulated quantum circuit as a whole. We adaptively divide a quantum circuit into clusters of entanglement according to the amount of memory available in our classical computing system, allowing the algorithm to exactly contract multiple gate layers at the same time. We call this variant the ``cluster-TEBD'' algorithm. For the latter, we devise a protocol, we term the ``Dynamical Adaptive Grouping Routine'', which makes use of hypergraph partitioning~\cite{Schlag2022KaHyPar} to automatically determine how to group qubits throughout the algorithm. We exploit this routine to adaptively vary the grouping in each iteration, based on the distribution of entangling gates in the contracted section of the circuit. Moreover, we employ heuristic approaches to contract multiple layers efficiently~\cite{Smith2018opt_einsum,Gray2021HyperExact}.

Benchmarking the original algorithms against their respective optimized variations on random-structured quantum circuits, with both Clifford and non-Clifford gates~\cite{Gottesman1998GKTheorem}, as well as on a tensor network-based implementation of Shor's algorithm for factoring~\cite{Shor1999}, shows substantial improvements in both runtime and fidelity, proving the effectiveness of cluster-based optimization in finite-fidelity tensor network algorithms. 

The paper is organized as follows. In Sec.~\ref{sec:qc-tn-review}, we make a comprehensive review of tensor network methods to simulate quantum circuits with finite fidelity. In Sec.~\ref{sec:clusters}, we explain how arranging qubits in clusters of highly entangled subsystems can improve TEBD and DMRG simulations of quantum circuits. In Sec.~\ref{sec:results}, we present the results of our benchmarks, where we compare runtime and fidelity of cluster-TEBD and standard TEBD when simulating random-structured quantum circuits and Shor's algorithm, and show the effect of contracting more layers in a single iteration of DMRG. Finally, in Sec.~\ref{sec:discussion}, we draw our conclusions and discuss possible future directions. 

\section{Simulating quantum circuits with tensor networks}
\label{sec:qc-tn-review}

\begin{figure}[t]
  \includegraphics[width=8.66cm]{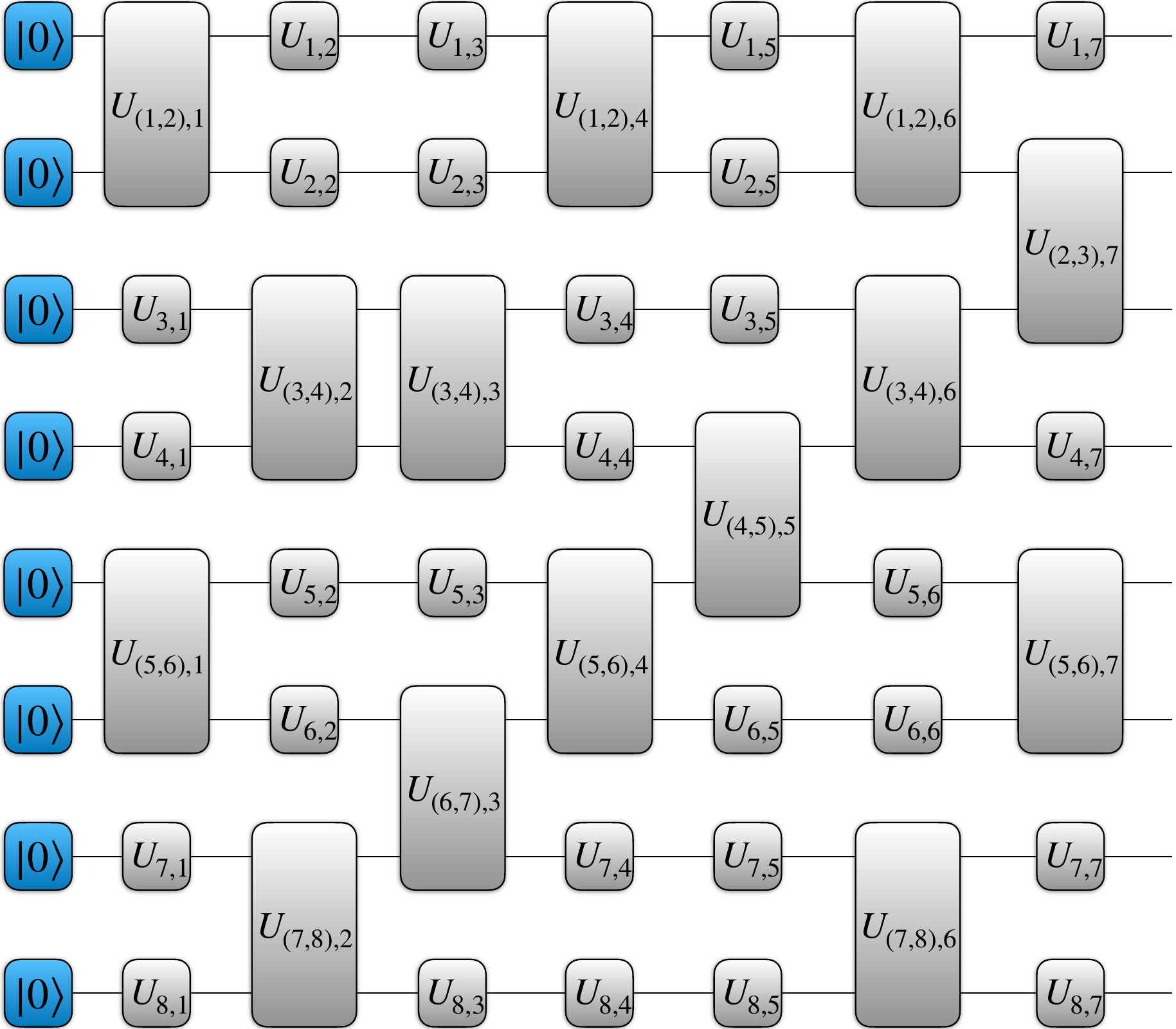}
  \caption{Quantum circuit with eight qubits and seven layers modeled as a tensor network. The initial state is an outer product of rank-1 tensors (vectors), as in this case it is a fully separable state. The gates are modeled as rank-2 tensors (matrices) for single-qubit operations, and rank-4 tensors for two-qubit operations, and they are labeled according to the qubits they act on and the layer they belong to.}
  \label{fig:circuit}
\end{figure}
In our computational framework, a tensor can be interpreted as a multidimensional array, with a number $r$ of indices, called the ``rank'' of the tensor, each one taking a finite number of values (``dimension''). Contracting tensors over their shared indices generates what is referred to as a ``tensor network''~\cite{Orus2014IntroTN,Montangero2018TNReview}. The purpose of this section is to introduce notation and briefly illustrate how tensor networks can be used to represent and simulate quantum circuits. In particular, we present some recent developments in the context of tensor network algorithms to simulate quantum circuits~\cite{Zhou2020WhatLimits, Ayral2023DMRG_QC}, introducing some quantities that will be crucial throughout the text.

\subsection{Mapping quantum circuits to tensor networks}
$N$-qubit gates in a quantum circuit can be directly represented as rank-$2N$ tensors with all indices of dimension 2. This means, for instance, that a two-qubit gate, which is usually represented as a $4 \times 4$ matrix, can be mapped to a $2 \times 2 \times 2 \times 2$ tensor. When no qubits are entangled at the start, the initial state of a circuit can also be directly represented as an outer product of rank-1 tensors (``vectors''), each of physical dimension $d = 2$, corresponding to the dimension of the Hilbert space of a qubit. 

When dealing with high-depth quantum circuits, it is crucial to define a quantity that characterizes the position of each gate within the overall evolution. For this reason, we define a gate layer $l$ in a quantum circuit as the subsets of gates that operate on independent sets of qubits at most once at the same time step. Let us identify as $U_{(q_1, \dots, q_n), l}$ a generic $n$-qubit gate applied on qubits $q_1, \dots, q_n$ at layer~$l$. For instance, the gates $U_{(3,4), l}$ and $U_{4, l'}$ cannot belong to the same layer, as they operate on the same qubit $4$, making $l \neq l'$. On the other hand, gates $U_{(3,4),l}$ and $U_{5, l''}$ could belong to the same layer, provided no operations are on qubit $5$ after $U_{(3,4),l}$. If this condition is satisfied, we can write $l = l''$. Gates belonging to the same layer can be physically interpreted as unitary operations applied at the same time step. Fig.~\ref{fig:circuit} shows an example of a quantum circuit modeled as a tensor network, using the described notation for layers. Because of the way a gate layer was defined, the number of gates per layer can change depending on the structure of the circuit. In the circuit shown in Fig.~\ref{fig:circuit}, all layers are fully filled with gates, but if, for instance, we added two gates $U_{(6,7)}$ and $U_{(7,8)}$ at the end of the evolution, each of them would belong to two new separate layers $8$ and $9$, respectively.

\begin{figure}[t]
  \includegraphics[width=8.66cm]{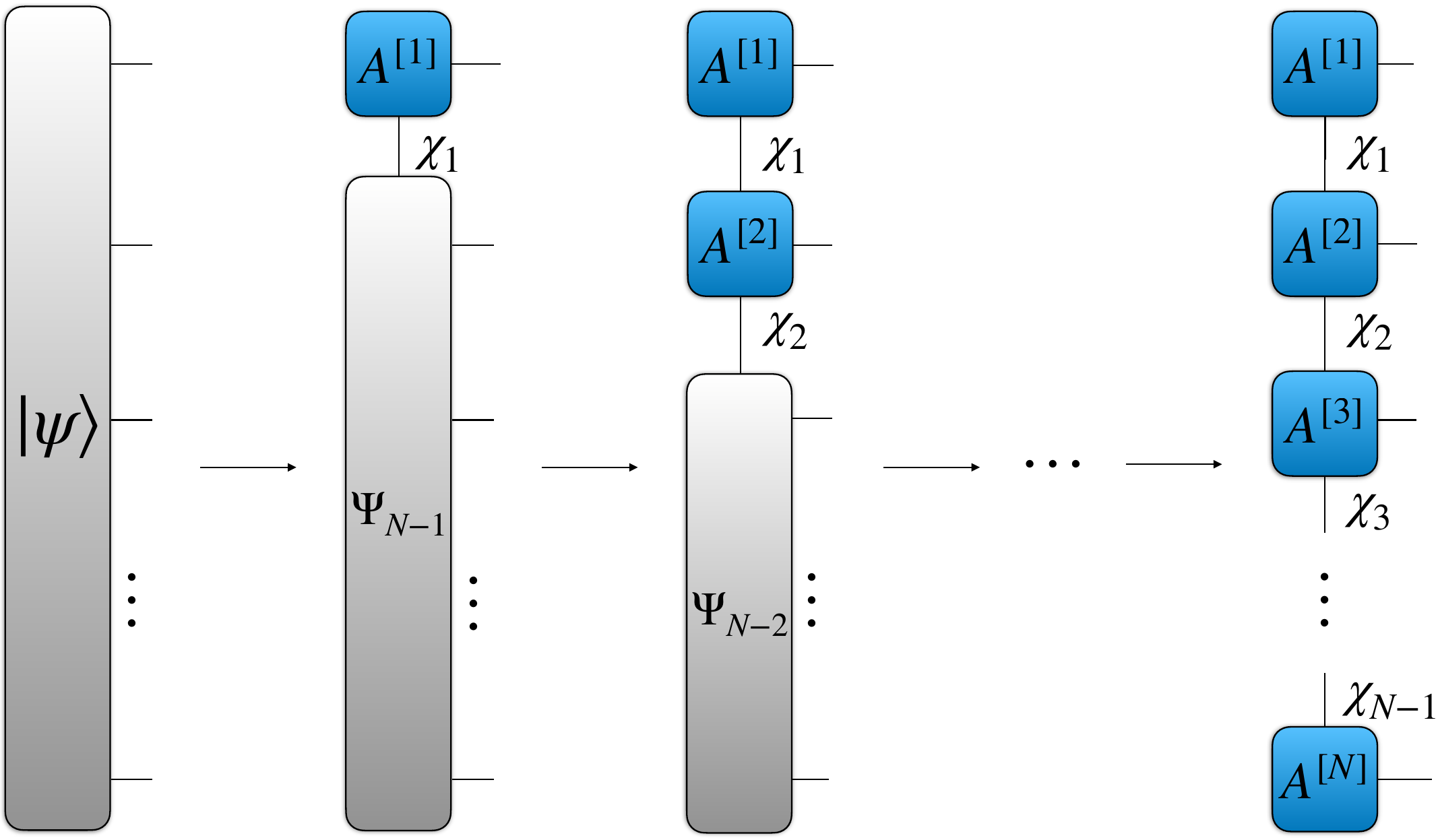}
  \caption{Entangled $N$-qubit quantum state $\ket{\psi}$, initially represented as a rank-$N$ tensor with total size $2^N$, decomposed into an $N$-site matrix-product state with successive singular value decompositions.}
    \label{fig:firstMPS}
\end{figure}

\subsection{Encoding entanglement: bond dimensions, truncation, and fidelity}

When the initial state of a quantum circuit is entangled, mapping it to tensor network notation is not as straightforward. Since the individual sites are not fully separable into tensor products, the entire state has to be stored in a  tensor of size $2^N$, with $N$ being the number of qubits. However, such a tensor can be decomposed, as shown in Fig.~\ref{fig:firstMPS}, by performing successive singular value decompositions (SVDs) to obtain an MPS representation~\cite{Fannes1992FirstMPS,Perez-Garcia2007MPS}. Aside from the physical dimensions, this representation introduces additional virtual bond dimensions $\chi_b$, which encode the entanglement of the state. Assuming for simplicity $\chi_1 = \chi_2 = \dots = \chi$, such a representation, in principle, contains a number of parameters equal to $2N\chi^2$, i.e., from exponential to linear in the number of qubits. Note that a separable state can be represented equivalently as an MPS with bond dimensions $\chi_b = 1 \ \forall \ b$.

When considering the bipartition of a quantum system related to a specific bond $b$, the entanglement between the two subsystems $A$ and $B$ can be quantified via the von~Neumann~entanglement~entropy
$S(\rho_A):=-\text{Tr}\left(\rho_A\log\rho_A\right)$, where $\rho_A:=\text{Tr}_B\left(\lvert \psi \rangle \langle \psi \rvert\right)$ is the reduced density matrix of subsystem $A$. We can equivalently denote the entanglement entropy as $S_b$, in relation to the bond $b$ where the bipartition originates from. It can be proven that the bond dimension $\chi_b$ of an MPS at bond $b$ scales exponentially with the bipartite entanglement entropy in the corresponding bond~\cite{Vidal2004EntScal, Schuch2008chi}:
\begin{equation}
    \chi_b \propto 2^{S_b}.
\end{equation}
Therefore, when simulating equilibrium states or unitary evolutions with MPS-based algorithms, it is crucial to estimate how the entanglement scales within the system. 

\begin{figure}[t]
  \includegraphics[width=8.66cm]{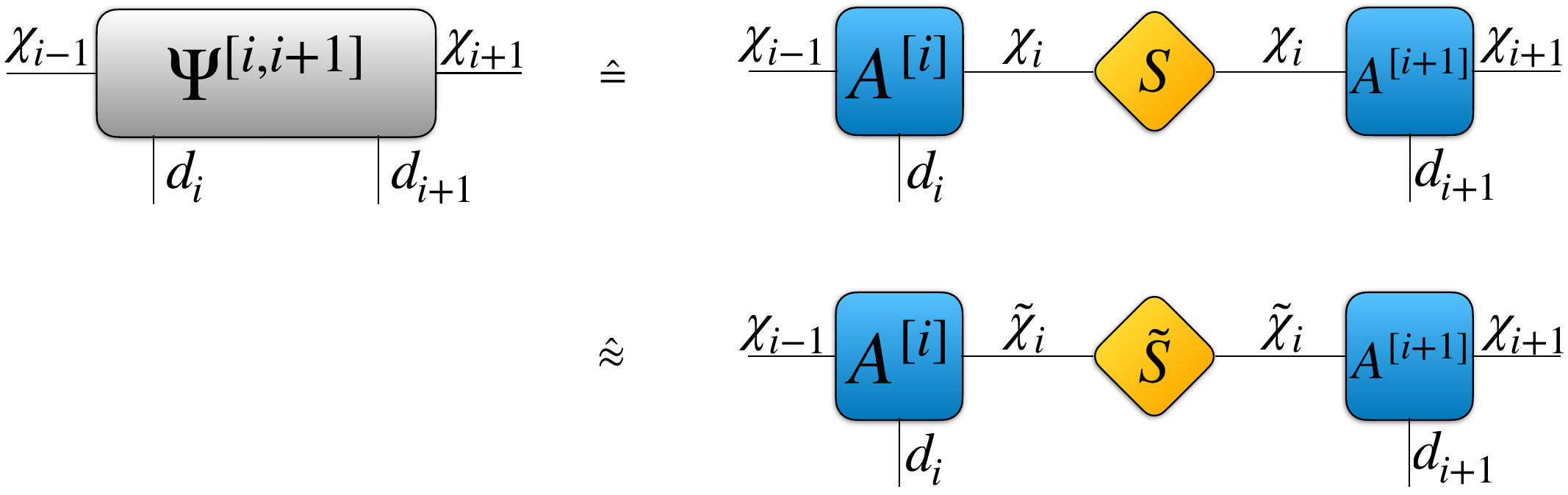}
  \caption{Truncating singular values in an SVD step. The matrix of singular values $S \in \mathbb{R}^{\chi_i \times \chi_i}$ is reduced to a matrix $\tilde{S} \in \mathbb{R}^{\tilde{\chi}_i \times \tilde{\chi}_i}$ with $\tilde{\chi} \leq \chi$.}
  \label{fig:trunc_svd}
\end{figure}

In general, unitary evolutions, as those performed in quantum circuits, can lead to volume-law scaling, where the entanglement entropy scales linearly with the system size, e.g., for a one-dimensional system of size $L$:
\begin{equation}
  S_b \propto L \implies \chi_b \propto 2^{L} .
\end{equation} 
In this case, the bond dimension scales exponentially with the system size. Inevitably, the longer the evolution, the larger the bond dimension, making exact simulations extremely challenging. An approximation can be introduced by reducing the bond dimension via the truncation of singular values in the SVD step, as in Fig.~\ref{fig:trunc_svd}. The singular values to truncate at each step can be determined using two parameters, namely, the cutoff~$\eta$ and the maximum bond dimension~$\chi_{\mathrm{max}}$. The cutoff~$\eta$ determines the minimum magnitude a singular value must have to be kept, meaning all singular values smaller than $\eta$ are discarded. The maximum bond dimension~$\chi_{\mathrm{max}}$ is the limit on the number of singular values to be kept, meaning if the bond dimension is larger than $\chi_{\mathrm{max}}$, all values in excess are discarded. 

To measure how distant a truncated state is from its exact counterpart, it is convenient to introduce the fidelity~$\mathcal{F}(n)$ of a quantum state after applying $n$~two-qubit gates. If the exact state is known, the fidelity can be measured as an inner product between the exact and the truncated MPS:
\begin{equation}
  \mathcal{F}(n) = \lvert\braket{\psi^{\mathrm{MPS}}_{\mathrm{exact}}(n) \vert \psi^{\mathrm{MPS}}_{\mathrm{trunc}}(n)}\rvert^2 .
\end{equation}
However, knowing the exact representation of a highly entangled quantum state requires large bond dimensions, leading back to the original problem. A way to circumvent this problem is to approximate the fidelity using singular values, as shown in Ref.~\cite{Zhou2020WhatLimits}. Let us define the two-qubit fidelity~$f_i$ as the ratio between the sum of singular values after and before truncation:
\begin{equation}
  f_i = \displaystyle \left(\sum_{\mu = 1}^{\tilde{\chi}} \tilde{S}^2_{\mu, \mu}\right) \bigg/ \left(\sum_{\mu = 1}^{\chi} S^2_{\mu, \mu}\right) ,
\end{equation}
with $S_{\mu, \mu}$ being the diagonal element of the matrix of singular values $S$ and $\tilde{\chi} \leq \chi$ being the truncation. It immediately follows that, if $\tilde{\chi} = \chi$ or $\displaystyle \sum_{\mu = \tilde{\chi}+1}^{\chi} S^2_{\mu, \mu} \approx 0$, then $f_i \simeq 1$. Finally, the total fidelity $\mathcal{F}(n)$ can be approximated accurately as the product of each individual two-qubit fidelity:
\begin{equation}
  \label{eq:approxF}
  \mathcal{F}(n) \approx \displaystyle \prod_{i=1}^{n} f_i .
\end{equation}
Together with runtime, the total fidelity can be used as a benchmark to evaluate the efficiency and the accuracy of classical algorithms for simulating quantum dynamics~\cite{Zhou2020WhatLimits}.

\subsection{Algorithms for simulating quantum circuits}
\label{subsec:algos}
The main algorithms for simulating quantum circuits, originally developed for simulating static and dynamical evolution of quantum many-body systems, exploit the approximate fidelity introduced in Eq.~\eqref{eq:approxF} to measure the simulation error on large/highly entangled quantum states~\cite{Zhou2020WhatLimits,Ayral2023DMRG_QC}.

\begin{figure}[t]
  \includegraphics[width=8.66cm]{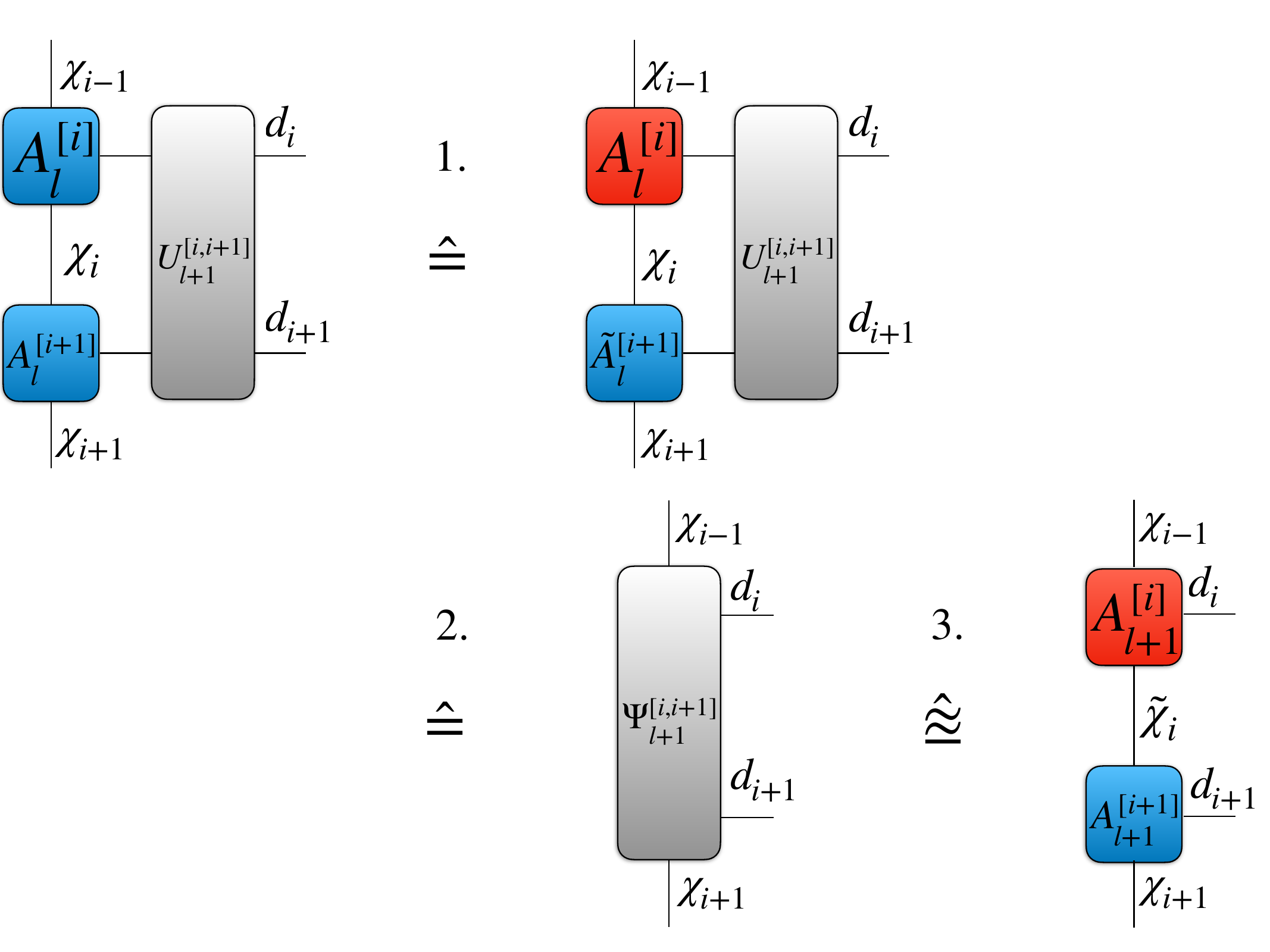}
  \caption{Single iteration of TEBD on a two-qubit gate. The red node symbolizes the orthogonality center of the MPS.}
  \label{fig:tebd_step}
\end{figure}

The TEBD algorithm iterates over each gate in the evolution and applies it to the corresponding qubits. Each iteration of the TEBD, analyzing a generic gate $U^{[i, i+1, \dots]}_{l+1}$, can be divided into three main steps:
\begin{enumerate}
  \item \textbf{Orthogonalization}: A sequence of QR decompositions is applied to the MPS tensors $A^{[j]}_l$ with $j~\neq~i$, bringing the state into a canonical form where all tensors to the left (right) of site $i$ satisfy left (right) isometry conditions. In this way, the norm of the state $\ket{\psi^{\mathrm{MPS}}}$ represented by the MPS can be computed as $\braket{\psi^{\mathrm{MPS}} \vert \psi^{\mathrm{MPS}}}~=~\text{Tr}\left(A^{[i]}_l{A^{[i]\dagger}_l}\right)$. We call the site $i$ the ``orthogonality center'' of the MPS.
  \item \textbf{Contraction}: The gate is applied to the initial state, obtaining tensor $\Psi^{[i, i+1,\dots]}_{l+1}$.
  \item \textbf{Truncated SVD}: Tensor $\Psi^{[i, i+1,\dots]}_{l+1}$ is decomposed back to MPS form, truncating its singular values according to parameters $\chi_{\mathrm{max}}$ and $\eta$. 
\end{enumerate}
The iteration is depicted in tensor diagram form in Fig.~\ref{fig:tebd_step}. Looking at the bigger picture of the full circuit and using the concept of layers, the TEBD essentially contracts quantum circuits one layer at a time, restoring the MPS after the application of each layer. 
The orthogonalization step constrains the order of operations but improves numerical stability and the fidelity of the simulation.

\begin{figure}[t]
  \includegraphics[width=8.66cm]{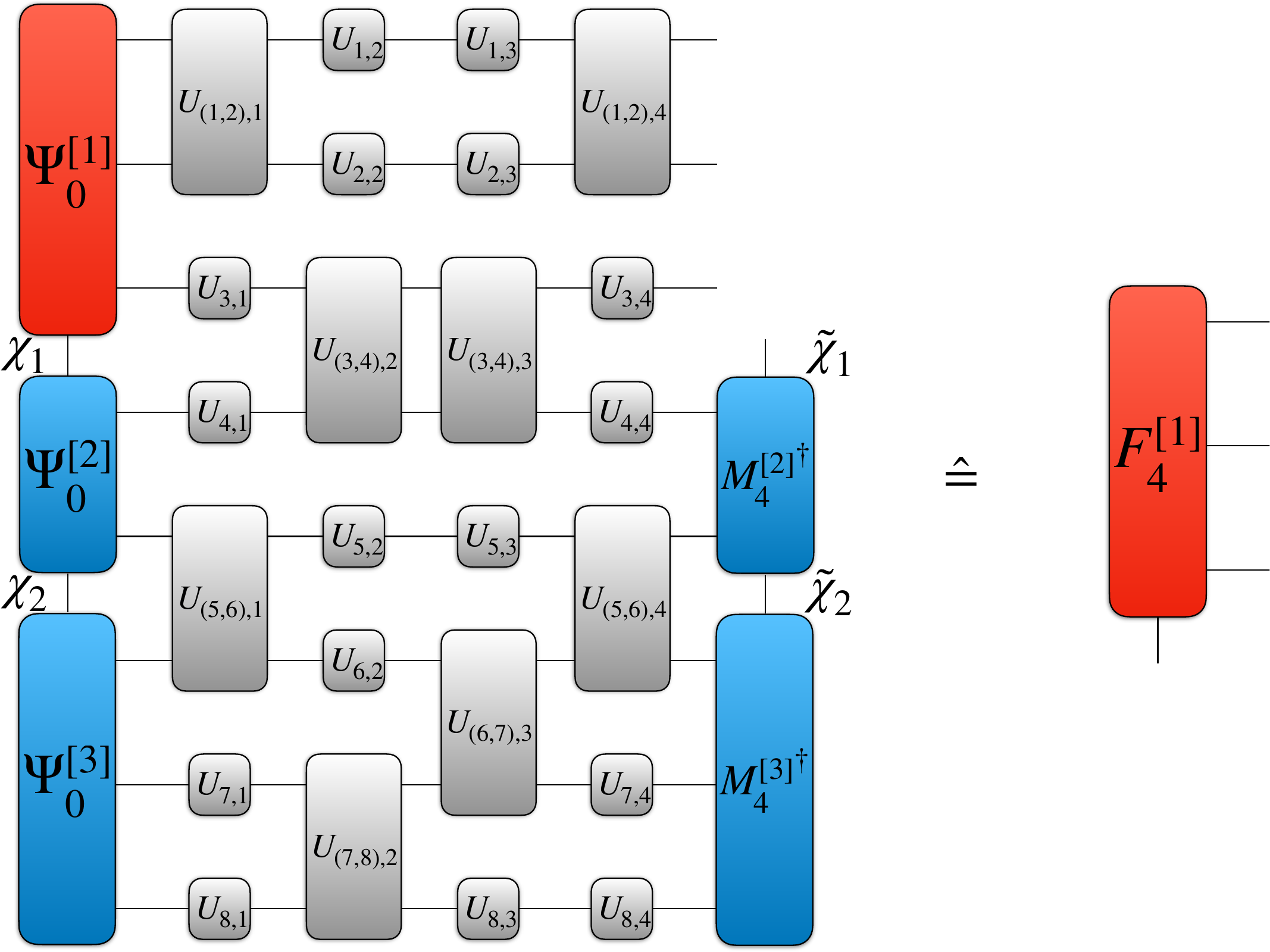}
  \caption{Contraction step of the DMRG algorithm for quantum circuits, setting $L_{\mathrm{max}} = 4$.}
  \label{fig:dmrg_qc_step}
\end{figure}

A different approach that requires the contraction of multiple layers of the circuit is the DMRG algorithm for quantum circuit simulations, introduced by Ayral \textit{et al.}~\cite{Ayral2023DMRG_QC}. The algorithm starts by defining a number of layers $L_{\mathrm{max}}$ to optimize in each step at the same time. Then, the initial MPS with tensors $A^{[1]}_0, \dots, A^{[N]}_0$ is mapped to a grouped MPS with tensors $\Psi^{[1]}_0, \dots, \Psi^{[g]}_0$, with $g \leq N$ the number of grouped sites. In a grouped MPS, each site $\tau$ has multiple physical indices $d_\tau$, depending on the number of qubits mapped to that site. Each grouped site is a tensor that represents a higher-dimensional system with a Hilbert space of dimension $2^{d_\tau}$, preserving all the entanglement within this group exactly without any truncation. A grouped site can be formed by contracting the sites from the original MPS that are to be combined, according to a predefined grouping scheme. The grouping scheme is predetermined prior to the algorithm’s execution and kept fixed throughout the simulation, and it is based on considerations about the structure of the circuits to simulate and on the amount of available memory. For instance, qubits on which many entangling operations are applied should belong to the same group, but the size of each grouped site should not exceed a certain threshold, as the system would run out of memory. In each step~$t$, the algorithm generates a random grouped MPS $M^{[1]\dag}_{t  L_{\mathrm{max}}}, \dots, M^{[g]\dag}_{t  L_{\mathrm{max}}}$ with bond dimensions $\tilde{\chi}_i$. Let $\Psi^{[1]}_{t  L_{\mathrm{max}}}, \dots, \Psi^{[g]}_{t  L_{\mathrm{max}}}$ denote the MPS of the evolved state after applying $t  L_{\mathrm{max}}$ layers of the circuit. The objective of the algorithm is to variationally minimize the distance between the random grouped MPS and the exact MPS representation of the state after $t  L_{\mathrm{max}}$ layers. For illustrative purposes, we describe the iteration of the algorithm at step $1$ in the following:
\begin{itemize}
  \item Orthogonalize $\Psi^{[1]}_{0}, \dots, \Psi^{[g]}_{L_{\mathrm{max}}}$ and $M^{[1]}_{L_{\mathrm{max}}},\dots,M^{[g]}_{L_{\mathrm{max}}}$, setting the orthogonality center at site $\tau$.
  \item Contract a tensor network composed of $\Psi^{[1]}_{0}, \dots, \Psi^{[g]}_{0}$ as initial state, gates from layer $1$ to layer $L_{\mathrm{max}}$ as evolution, and tensors $M^{[1]\dag}_{L_{\mathrm{max}}}, \dots, M^{[\tau-1]\dag}_{L_{\mathrm{max}}}, M^{[\tau+1]\dag}_{L_{\mathrm{max}}}, \dots, M^{[g]\dag}_{L_{\mathrm{max}}}$ applied at the end. Fig.~\ref{fig:dmrg_qc_step} shows this step for $L_{\mathrm{max}} = 4$. Denote the tensor resulting from the contraction of this network as $F^{[\tau]}_{L_{\mathrm{max}}}$.
  \item Compute $f_\tau=\text{Tr}\left(F^{[\tau]}_{L_{\mathrm{max}}}F^{[\tau]\dag}_{L_{\mathrm{max}}}\right)$ and replace $M^{[\tau]}_{L_{\mathrm{max}}}=\dfrac{F^{[\tau]}_{L_{\mathrm{max}}}}{f_\tau}$.
\end{itemize}
This iteration can be generalized for any step $t$, where the evolution from layer $(t-1)L_{\mathrm{max}}+1$ to layer $t L_{\mathrm{max}}$ is optimized (for further details, see Ref.~\cite{Ayral2023DMRG_QC}). 

In all iterations, each bond dimension $\tilde{\chi}_1, \dots, \tilde{\chi}_{g-1}$ belonging to the grouped MPS $M^{[1]\dag}_{t L_{\mathrm{max}}}, \dots, M^{[g]\dag}_{t L_{\mathrm{max}}}$ should be chosen large enough to encode the entanglement generated in its respective bond. An approximation can be introduced by setting a maximum bond dimension $\chi_{\mathrm{max}}$, such that if the amount of entanglement in any bond exceeds this threshold, the bond dimension is set to $\chi_{\mathrm{max}}$ and some information is lost.

A repetition of the described iteration for each grouped site~$\tau$ is called a ``sweep''. The quantity~$f_\tau$ is monotonically increasing and it represents the fidelity of $M^{[\tau]}_{t L_{\mathrm{max}}}$ with respect to the state $\Psi^{[\tau]}_{t L_{\mathrm{max}}}$. After a full sweep, assuming all bond dimensions $\tilde{\chi}_{\tau-1}, \tilde{\chi}_{\tau}$ are chosen large enough, eventually $M^{[\tau]}_{t L_{\mathrm{max}}} \simeq \Psi^{[\tau]}_{t L_{\mathrm{max}}}$ for every~$\tau$. When the amount of entanglement is larger than the maximum bond dimension $\chi_{\mathrm{max}}$ at any point throughout the simulation, it is also possible to repeat a full sweep multiple times to improve the fidelity of the final state.

\section{Entanglement clustering in finite-fidelity simulations}
\label{sec:clusters}

The core idea of this work is to improve the algorithms for simulating quantum circuits described in the previous section using cluster-based protocols. By leveraging the properties of a given circuit's structure, we can ``cluster'' qubits based on regions of the circuit with a large number of entangling gates. Previous work has proposed strategies that group gates into clusters~\cite{Fatima2021GateCluster, Chen2021CircuitPartitioner, cicero2024simulationquantumcomputersreview}. In our approach, which we refer to as ``entanglement clustering'', clusters are formed by grouping gates applied up to a certain layer on a subset of the total system. Gates within a cluster are not contracted immediately; instead, they are combined with the corresponding portion of the initial state into a tensor network, which is contracted exactly once the cluster reaches its maximal size. The termination criterion for each cluster depends on the size of the resulting tensor, accounting both for the arrangement of multiqubit gates in the analyzed circuit portion and the bond dimension of the state at that stage. This physics-informed clustering approach allows TEBD to adaptively divide the circuit into multilayered clusters, employing optimized tensor network contractions~\cite{Gray2021HyperExact} to contract each cluster exactly. In DMRG, a similar approach is applied through an adaptive clustering algorithm, which exploits hypergraph partitioning~\cite{Schlag2022KaHyPar} to automatically form clusters of qubits, based on the distribution of entangling operations within a specified segment of the circuit of up to $L_{\mathrm{max}}$ gate layers. This qubit grouping is used to generate a grouped MPS, whose sites are variationally optimized in each iteration. Our implementation of the DMRG algorithm is adaptive, as the structure of the grouped MPS dynamically changes at each iteration to match the entangling gate distribution in the local circuit region.

In this section, we introduce our cluster-based variations of the TEBD and DMRG algorithms.

\begin{figure*}[ht!]
  \begin{subfigure}{0.45\textwidth}
      \caption{}
      \label{fig:cTEBD_example_a}
      \centering
      \includegraphics[width=8.4cm]{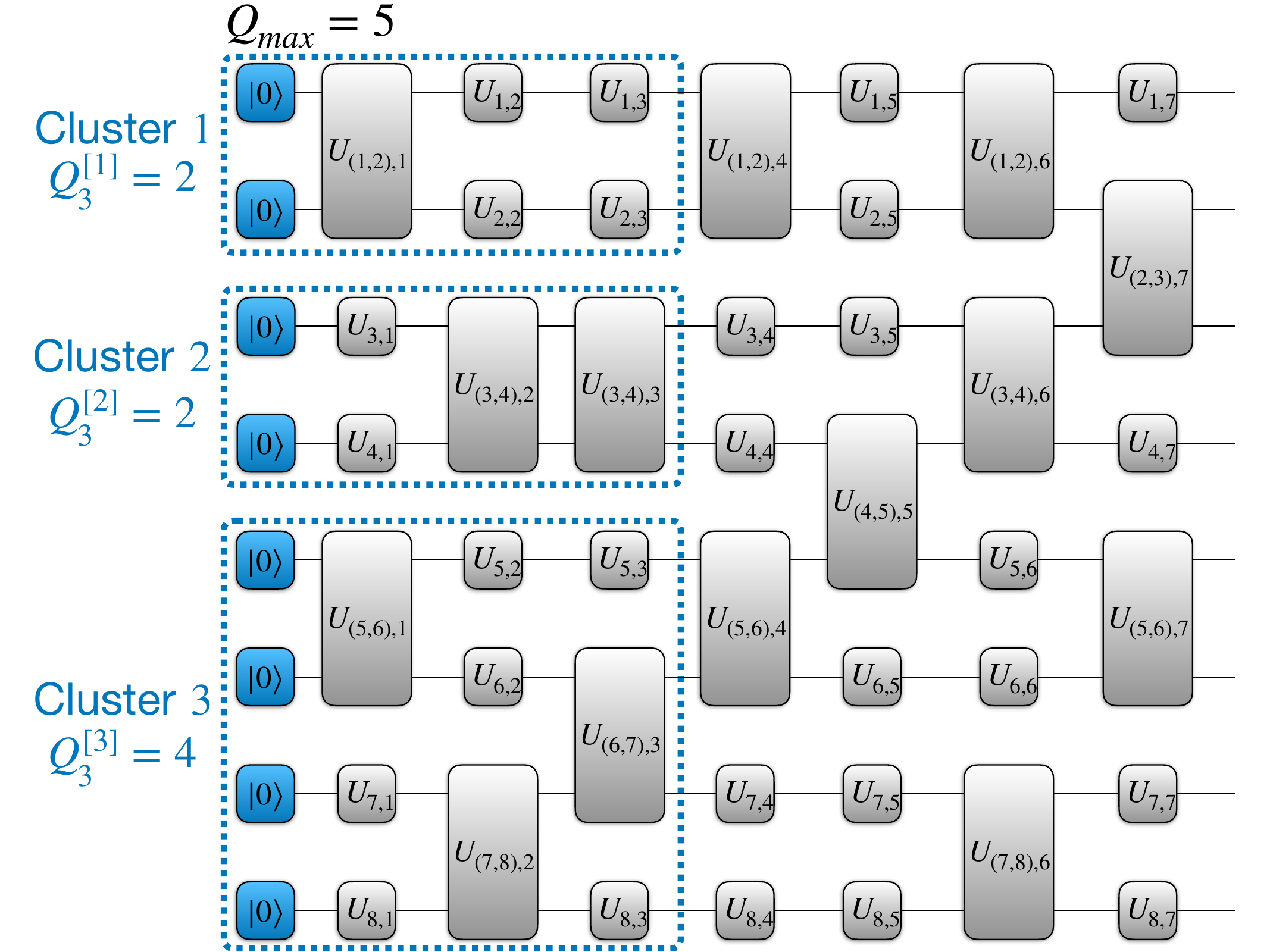}
  \end{subfigure}
  \hspace{10mm}
  \begin{subfigure}{0.45\textwidth}
      \caption{}
      \label{fig:cTEBD_example_b}
      \centering
      \includegraphics[width=8.4cm]{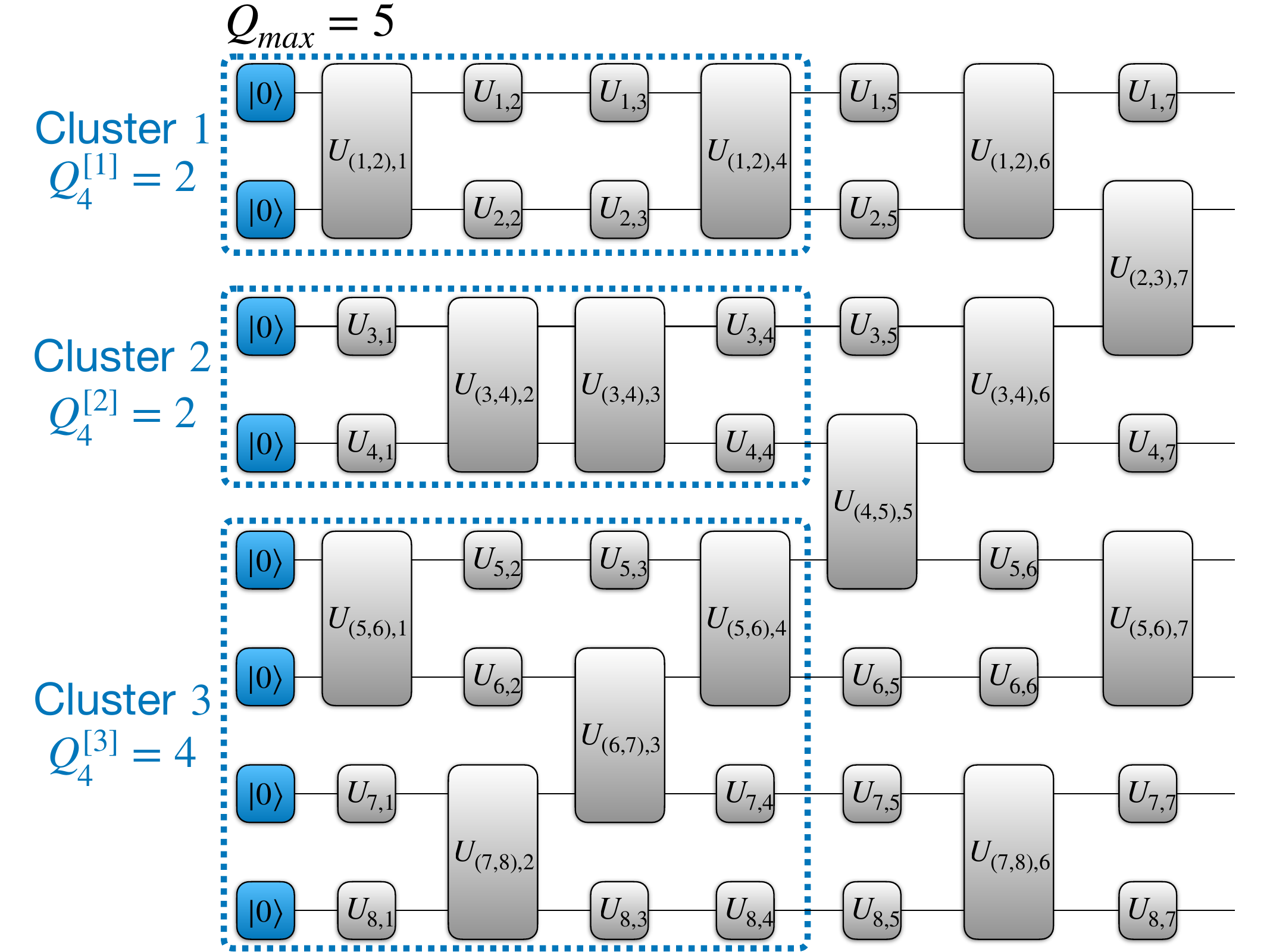}
  \end{subfigure}
  \hspace{1mm}
  \begin{subfigure}{0.45\textwidth}
      \caption{}
      \label{fig:cTEBD_example_c}
      \centering
      \includegraphics[width=8.4cm]{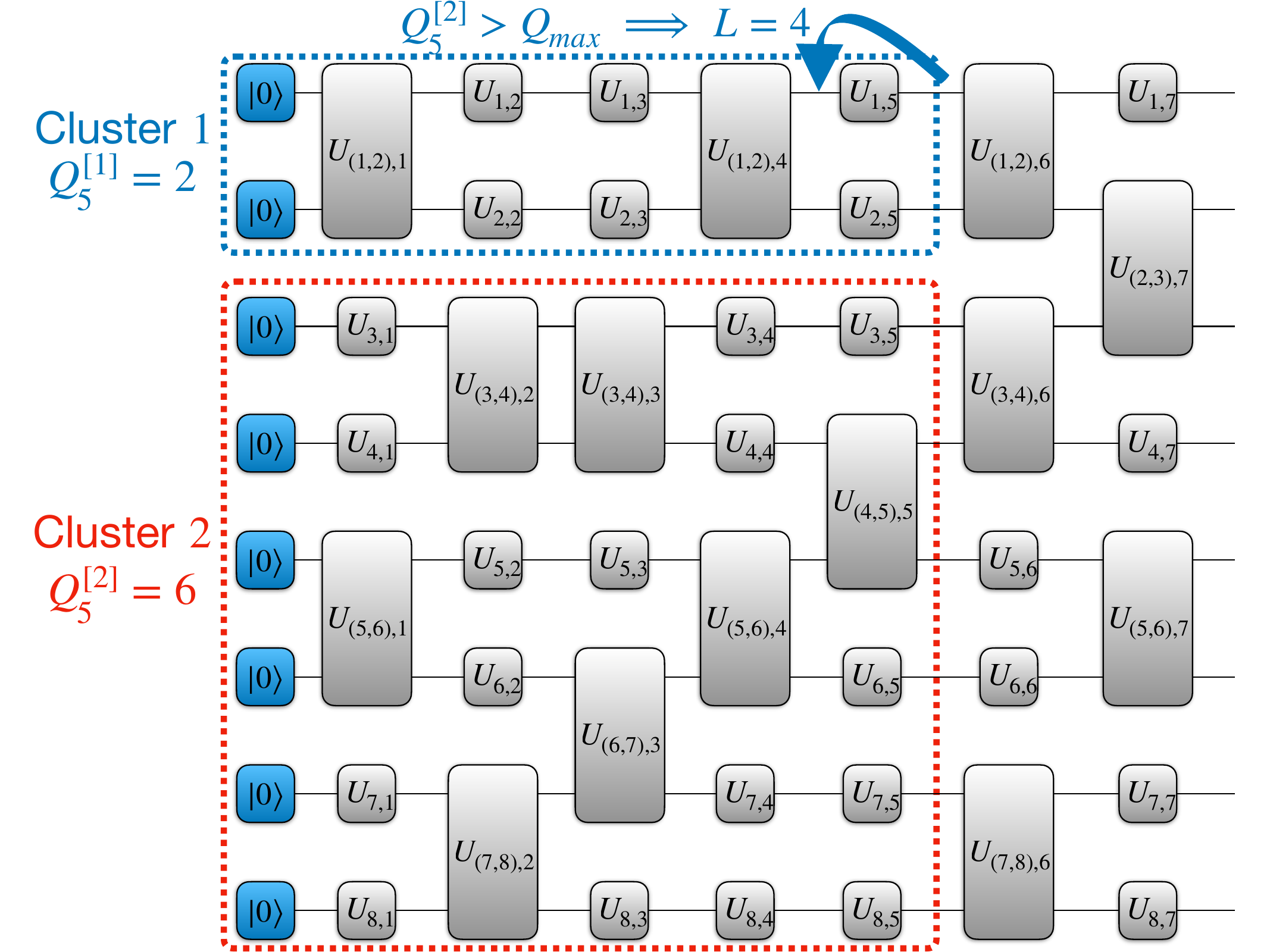}
  \end{subfigure}
  \hspace{10mm}
  \begin{subfigure}{0.45\textwidth}
      \caption{}
      \label{fig:cTEBD_example_d}
      \centering
      \includegraphics[width=8.4cm]{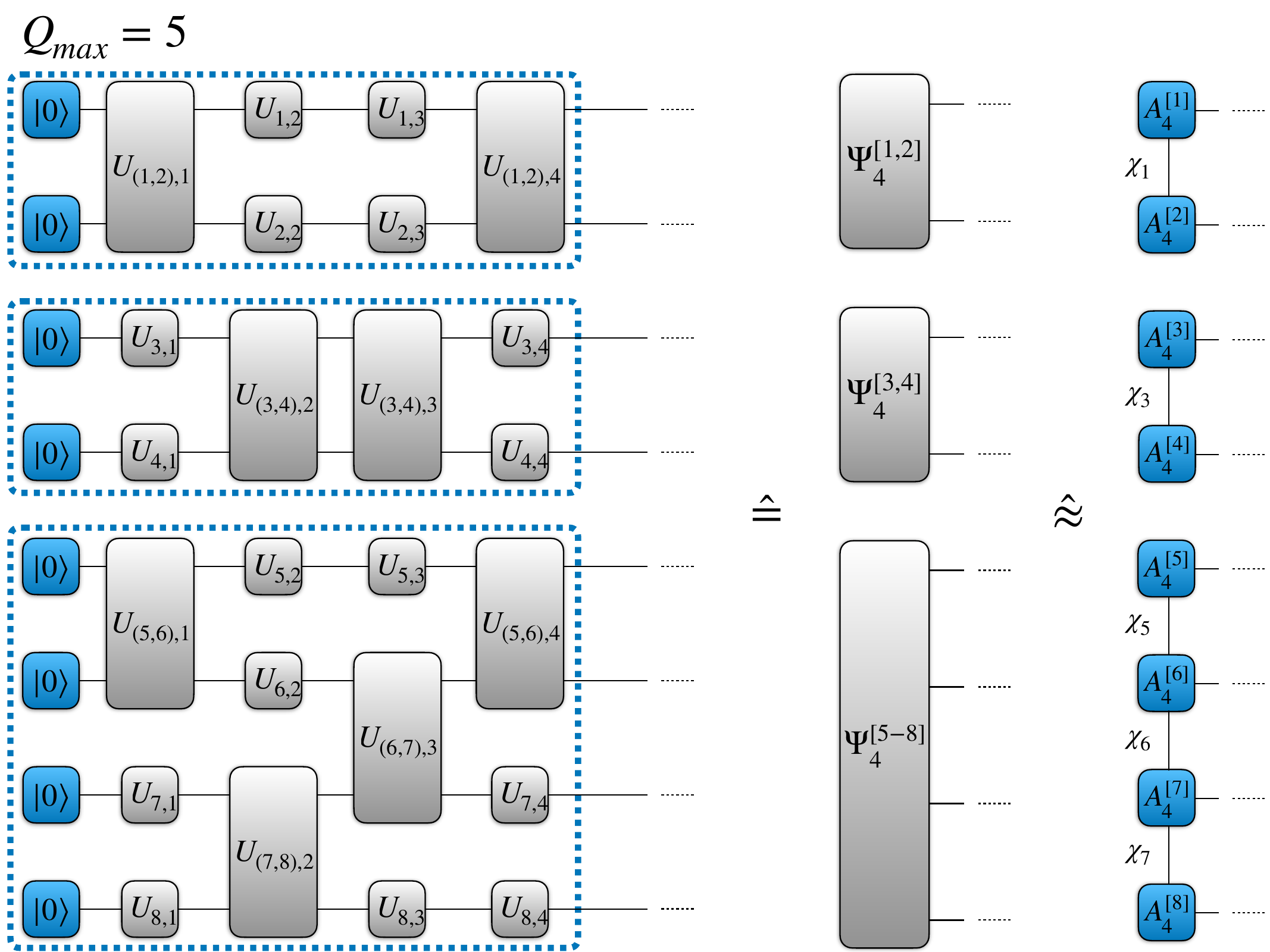}
  \end{subfigure}
  \caption{Schematic of the iteration process of the cluster-TEBD on an eight-qubit, seven-layer quantum circuit, with $Q_{\mathrm{max}} = 5$. \hyperref[fig:cTEBD_example_a]{(a)}~Up to layer $2$, the circuit forms four clusters of entanglement, with two qubits each. At layer $3$, the operation $U_{(6,7),3}$ groups clusters $3$ and $4$ in a new cluster $3$ with four qubits. As the size of this cluster does not violate the condition on $Q_{\mathrm{max}}$, we move on to the next layer. \hyperref[fig:cTEBD_example_b]{(b)}~At layer $4$, operations $U_{(1,2),4}$ and $U_{(5,6),4}$ are performed on pairs of qubits that are already in the same cluster. Consequently, the clusters remain the same and we move on to the next layer. \hyperref[fig:cTEBD_example_c]{(c)}~At layer $5$, the operation $U_{(4,5),5}$ groups clusters $2$ and $3$ into a new cluster $2$ with six qubits. As the size of this cluster violates the condition on $Q_{\mathrm{max}}$, layer $4$ is the chosen layer up to which the circuit can be contracted exactly. \hyperref[fig:cTEBD_example_d]{(d)}~Schematic of the division of clusters up to layer $4$, contraction and decomposition -- with truncation, if necessary. The obtained MPS is used as the initial state of the remaining evolution.}
    \label{fig:cTEBD_example}
\end{figure*}

\subsection{Cluster-TEBD}
\label{subsec:ctebd}

The cluster-TEBD algorithm modifies the TEBD algorithm described in Sec.~\ref{subsec:algos} to allow the exact contraction of multiple gate layers of a quantum circuit. The idea is to divide the network into clusters of entanglement, based on the structure of the circuit to simulate. These clusters are independent from each other, meaning they can be interpreted as intermediate-to-large-sized tensor networks that can be exactly contracted in an efficient way. 

The idea of gate clustering has already been successfully applied in the context of exact simulations of quantum circuits~\cite{Fatima2021GateCluster, Chen2021CircuitPartitioner, cicero2024simulationquantumcomputersreview}, and our purpose is to exploit it in finite-fidelity tensor network algorithms. 

In the context of exact tensor network contractions, a similar approach is the ``slicing'' technique~\cite{Gray2021HyperExact, Chen2018classicalsimulationintermediatesizequantum, Villalonga2019hpcsimqc, Huang2020classical, Schutski2020SimpleHeuristics, zhang2019alibabacloudquantumdevelopment, Huang2021Slicing, cicero2024simulationquantumcomputersreview, tang2025tensorqcscalabledistributedquantum, herzog2025jointcuttinghybridschrodingerfeynman,sarode2025circuitpartitioningcircuitexecution}, where a network is divided into independent subnetworks by choosing a subset of indices to fix. As a consequence, the contraction of all subnetworks has to be repeated multiple times, for every possible value of the fixed indices. The main difference in our approach is that no indices within the network are fixed, eliminating the need to repeat the contraction of the generated subnetworks. Instead, the circuit partitioning is only based on the entanglement distribution and on the available memory, making sure that, after all subnetworks have been contracted, the resulting tensor sizes are kept manageable for the SVD compression step.

A single iteration of cluster-TEBD is performed as follows. We define an array of length~$N-1$, representing bonds between neighboring qubits, with all entries initially set to \texttt{False}. Each entry $i$ in the array corresponds to the bond between qubits $i$ and $i+1$, and its value indicates whether an entangling operation has been applied between them. The algorithm iterates over all gates in the circuit, starting from the layer closest to the initial state, to track the distribution of entangling operations. When encountering an $n$-qubit gate, with $n > 1$, acting on qubits $i, \dots, i+n-1$, the corresponding entries $i, \dots, i+n-2$ of the bond array are updated to \texttt{True}. This signifies that an entangling operation has been applied between these qubits. At the end of each layer, the algorithm identifies entanglement clusters by finding consecutive entries in the bond array that are set to \texttt{True}. In the context of the cluster-TEBD algorithm, an entanglement cluster can be interpreted as a portion of the circuit, up to the current layer $L_{\text{current}}$, where qubits frequently interact with each other but never interact with the rest of the system. Since distinct clusters of entanglement can be treated independently, the corresponding circuit portion can be contracted as a separate circuit. The size of the tensor resulting from the contraction of a cluster depends on two quantities:

\begin{enumerate}
  \item the size of a cluster up to $L_{\text{current}}$, i.e., the number of consecutive entries of the bond array equal to \texttt{True}. The size of cluster $k$ can be expressed as $q_n^{[k]}-q_1^{[k]}+2$, with $q_1^{[k]}, \dots, q_n^{[k]}$ being the qubits belonging to cluster $k$ such that $\mathtt{bond[q_i^{[k]}] = True}$ for $i \in 1, \dots, n$;
  
  \item the bond dimensions of the initial state $\chi_{q_1^{[k]}-1}, \chi_{q_n^{[k]}}$ that are adjacent to cluster $k$.
\end{enumerate}
Knowing that all the physical indices of a quantum circuit  have dimension $2$, the size of tensor $\Psi^{[k]}_{L_\text{current}}$ that exactly encodes the entanglement cluster $k$ up to layer $L_{\text{current}}$ can be computed as
\begin{equation}
    \text{size}\left(\Psi^{[k]}_{L_{\text{current}}}\right) = 2^{q_n^{[k]}-q_1^{[k]}+2}\chi_{q_1^{[k]}-1}\chi_{q_n^{[k]}} .
\end{equation}
For the clustering mechanism to work correctly, once all gates in layer $L_{\text{current}}$ have been analyzed, the algorithm needs to ensure that, for every $k$, tensors $\Psi^{[k]}_{L_{\text{current}}}$ resulting from the exact contraction of cluster $k$ can be stored in memory. To do this, the algorithm uses an input parameter $Q_{\mathrm{max}}$ and checks that
\begin{equation}
\log_2\left[\text{size}\left(\Psi^{[k]}_{L_{\text{current}}}\right)\right] \leq Q_{\mathrm{max}} \, , \quad \forall \ k .
    \label{eq:cond_cTEBD}
\end{equation}
The condition above should guarantee that the clusters do not exceed the system's memory capacity. For quantum circuits with a large number of layers and few gates per layer, it is useful to impose an additional constraint on the maximum number of layers~$L_{\mathrm{max}}$ that can be contracted in one step, as follows:
\begin{equation}
  \label{eq:cond2_cTEBD}
  L_{\text{current}} \leq L_{\mathrm{max}} .
\end{equation}
The quantum circuit can then be contracted exactly up to the last layer~$L$ that satisfies both Eqs.~\eqref{eq:cond_cTEBD} and~\eqref{eq:cond2_cTEBD}. Once layer~$L$ has been identified, the algorithm proceeds to exactly contract all entanglement clusters, dividing each portion of the circuit from the initial state up to~$L$ into independent tensor networks. The exact contraction of each cluster~$k$ exploits the greedy algorithm to find the corresponding contraction sequences~\cite{Smith2018opt_einsum,Gray2021HyperExact}. The resulting tensors $\Psi_{L}^{[k]}~\in~\mathbb{C}^{\chi_{q_1^{[k]}-1} \times q_1^{[k]} \times \dots \times q_n^{[k]} \times \chi_{q_n^{[k]}}}$ are finally decomposed using successive SVDs -- as shown back in Fig.~\ref{fig:firstMPS}. 

Once all clusters have been contracted and decomposed, we obtain a new initial  MPS $A_{L}^{[1]}, \dots, A_{L}^{[N]}$, and the iteration is repeated on the remaining evolution, with the new initial state, until all gates have been contracted. Since the updated initial state is restored in its MPS form at the end of each iteration, the clusters are also reset for subsequent iterations. This enables the adaptive formation of new clusters, which may differ from those in previous iterations, based on the structure of the remaining circuit and the bond dimensions of the new initial state.  Fig.~\ref{fig:cTEBD_example} depicts the first iteration of the cluster-TEBD algorithm applied to the circuit from Fig.~\ref{fig:circuit}, setting $Q_{\mathrm{max}} = 5$.

\begin{figure*}[t!]
    \begin{subfigure}{0.44\textwidth}
      \caption{}
      \label{fig:dmrg_grouping_a}
      \centering
      \includegraphics[width=7.6cm]{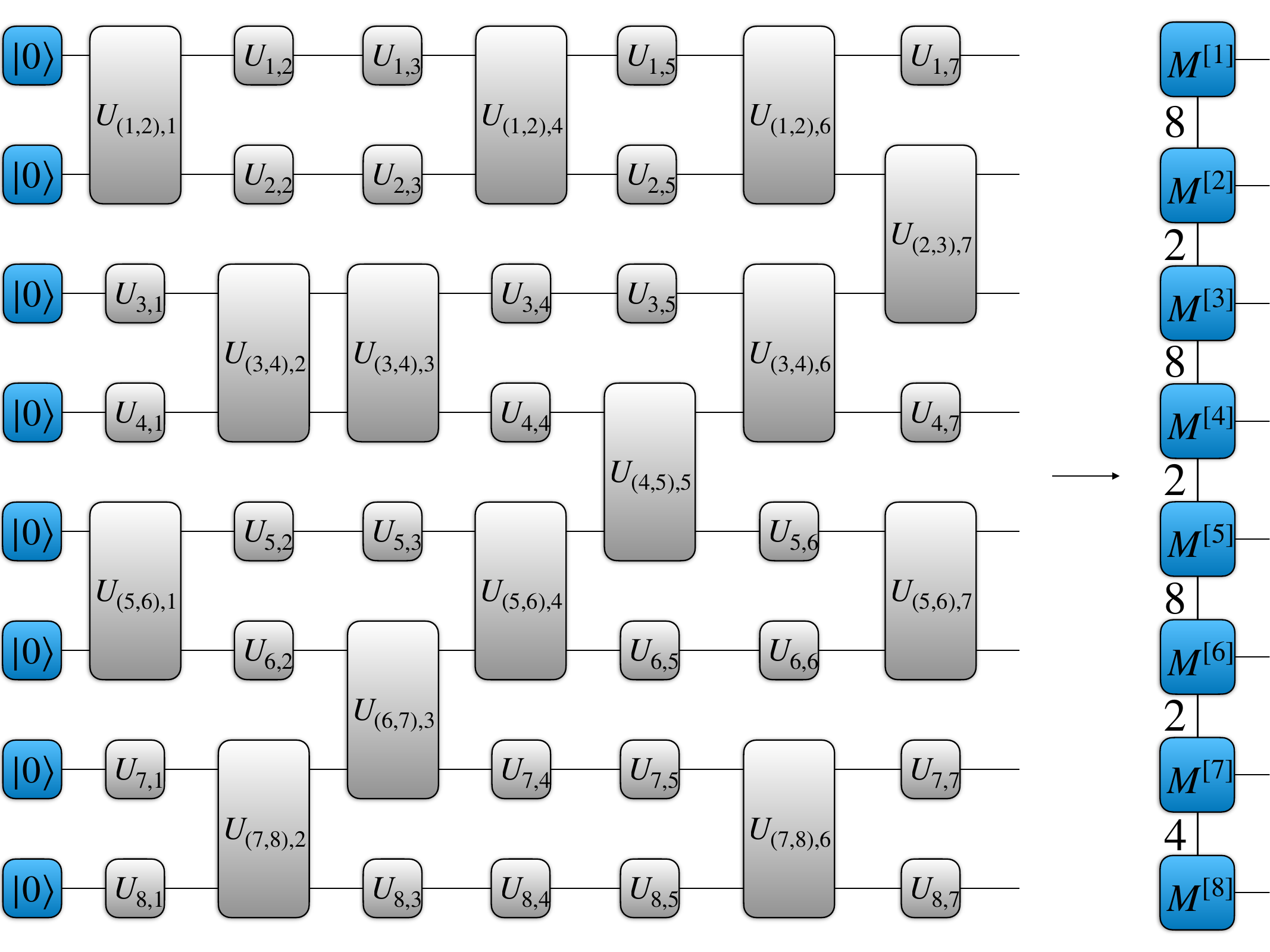}
  \end{subfigure}
  \hspace{10mm}
  \begin{subfigure}{0.48\textwidth}
      \caption{}
      \label{fig:dmrg_grouping_b}
      \centering
      \includegraphics[width=8.6cm]{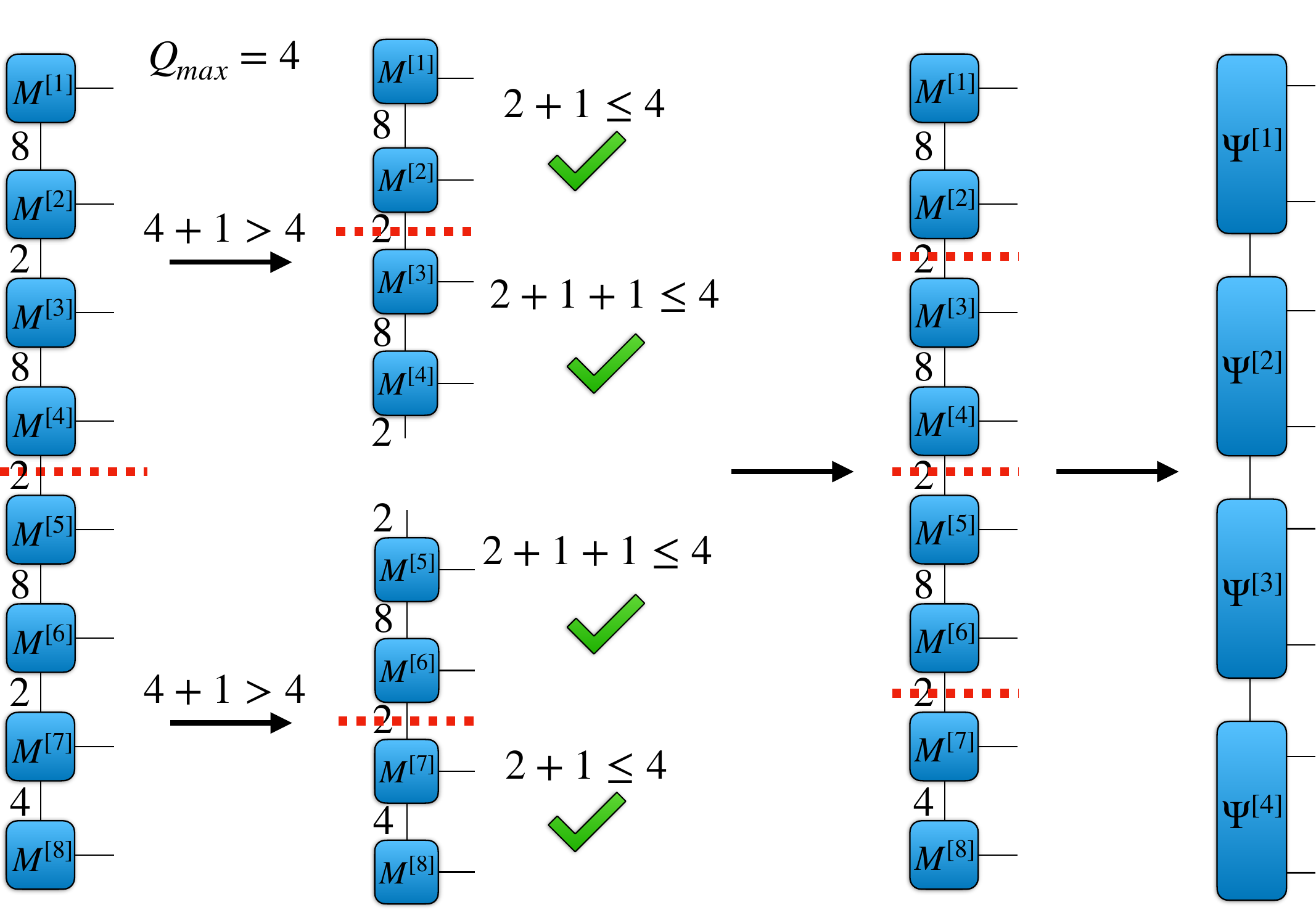}
  \end{subfigure}
  \caption{Dynamical adaptive grouping routine for DMRG on an eight-qubit quantum circuit, setting $L_{\mathrm{max}} = 7$. \hyperref[fig:dmrg_grouping_a]{(a)}~Generation of the virtual MPS from the quantum circuit by counting the number of two-qubit gates per bond and estimating the bond dimensions. \hyperref[fig:dmrg_grouping_b]{(b)}~Recursive bipartitioning protocol for the virtual MPS, setting $Q_{\mathrm{max}} = 4$. The generated partitioning is finally used to group the initial state accordingly, creating the grouped MPS for the DMRG algorithm.}
    \label{fig:dmrg_grouping}
\end{figure*}

\subsection{Dynamical Adaptive Grouping Routine for DMRG}
\label{subsec:hdmrg}
The idea of clustering together qubits based on the distribution of entanglement in a quantum circuit can also be used to improve the DMRG algorithm from Sec.~\ref{subsec:algos}. As opposed to regular, periodically invariant circuit structures, for example, a brick-wall structure with alternating two-qubit gates, generic circuits with irregular structures display an uneven distribution of entanglement throughout the evolution. Depending on the time step, distinct subsets of qubits may experience numerous entangling operations, while other separate subsets may undergo very few or even no operations. Our goal is to develop entanglement-based protocols that enhance the applicability of DMRG for simulating generic quantum circuits.

In Sec.~\ref{subsec:algos}, we introduced the concept of a grouping scheme, designed to group together highly entangled sites in the simulated quantum circuit. This idea of forming a grouped MPS is conceptually similar to creating clusters of entanglement, with one key difference: in the original DMRG algorithm, the grouping was predetermined \emph{prior to} the algorithm's execution. While this method works well for circuits with regular structures, it becomes less effective for deep circuits with irregularly arranged entangling gates, where it is not straightforward to predict how the bond dimensions will grow in different portions of the circuit. Therefore, we introduce a clustering approach to adaptively choose the site grouping and the needed bond dimensions, according to the structure of the simulated circuit. 

An iteration of the dynamical adaptive grouping routine works as follows. Given a generic $N$-qubit quantum circuit, we count the number of entangling operations $E_{L_{\mathrm{max}}}(b)$ for each bond~$b$, up to the chosen number of contracted layers per iteration $L_{\mathrm{max}}$. Then, we create a virtual MPS with $N$ sites, where each bond dimension is defined as 
\begin{equation}
  \tilde{\chi}_b := \min\left\{2^{E_{L_{\mathrm{max}}}(b)}\chi_b, \, \chi_{\mathrm{max}}^{\mathrm{DMRG}}, \,2^{\min\left\{b, N - b\right\}}\right\} .
  \label{eq:chi_est}
\end{equation}
The quantity $2^{\min\left\{b, N - b\right\}}$ corresponds to the highest value the bond dimension can attain at bond~$b$, $\chi_b$~is the bond dimension of the initial state at bond~$b$, and $\chi_{\mathrm{max}}^{\mathrm{DMRG}}$ is an input parameter of the algorithm, serving as a truncation parameter, similar to TEBD. In practice, by counting the number of entangling gates on each bond, we obtain an upper bound on the amount of entanglement that can be generated by the circuit on each bond. 

Since the virtual MPS provides direct information on the distribution of entanglement within the simulated quantum circuit section, we can use this structure to construct a grouped MPS by clustering qubits that are highly entangled throughout the evolution. The goal is to group together sites connected by larger bond dimensions, while using the lower ones as the effective bond dimensions of the grouped MPS. At the same time, we should verify that the generated grouped sites actually fit within the available memory. This problem can be solved by mapping the virtual MPS into a graph, where the bond dimensions are represented as edge weights. The task then becomes to find a \textit{min-cut} multipartitioning of the graph, such that qubits connected by large bond dimensions are grouped within the same partition, while the lower bond dimensions correspond to the edges cut by the partitioning process. To achieve this, we make use of the \texttt{KaHyPar} graph partitioning tool~\cite{Schlag2022KaHyPar} to perform recursive bipartitioning. In the first iteration, we divide the full virtual MPS into two subnetworks, with sites from $1$ to $k_1$ and from $k_1+1$ to $N$, respectively, and check the following condition for each subnetwork:
\begin{equation}
  N_1(p) + \log_2\chi_{k_1} \leq Q_{\mathrm{max}} ,
  \label{eq:cond_group1}
\end{equation}
with $N_1(p)$ being the number of qubits in partition $p$, $\chi_{k}$ the bond dimension that separates the two subnetworks, and $Q_{\mathrm{max}}$ a parameter that limits the size of each grouped site -- similar to cluster-TEBD. The number of qubits $N_1(p)$ in partition~$p$ can be easily estimated as $k_1$ for the first partition, and $N - k_1 + 1$ for the second one. If the condition in Eq.~\eqref{eq:cond_group1} does not hold for any of the two subnetworks, the algorithm recurs by applying the same procedure on the corresponding subnetwork. At a generic step $i > 1$ of the recursion, the previous condition can be generalized as follows:
\begin{equation}
  N_i(p) + \log_2\chi_{k_j} + \log_2\chi_{k_i} \leq Q_{\mathrm{max}},
  \label{eq:cond_group2}
\end{equation}
with $\chi_{k_j}$ being the bond dimension connecting two subnetworks that were separated at a previous recursion step~$j~<~i$. The algorithm stops recurring when Eq.~\eqref{eq:cond_group2} -- or Eq.~\eqref{eq:cond_group1} for subnetworks at the extremes of the MPS -- holds for all the generated subnetworks. Finally, we use the obtained partitioning of the virtual MPS as our grouping scheme. The sites of the initial state belonging to the same partition~$\tau$ are grouped in the corresponding site~$\tau$, forming the grouped MPS that will be used in the iteration. Fig.~\ref{fig:dmrg_grouping} depicts an example of the dynamical adaptive grouping routine applied on the circuit from Fig.~\ref{fig:circuit}. 

Once the grouped MPS is defined, the algorithm can proceed with the iteration, employing the qubit grouping selected with the adaptive clustering algorithm, and using $\tilde{\chi}_b$ from Eq.~\eqref{eq:chi_est} as an estimate for the bond dimensions of the grouped MPS $M^{[1]\dag}_{t  L_{\mathrm{max}}}, \dots, M^{[g]\dag}_{t  L_{\mathrm{max}}}$ applied at the end of the evolution. Finally, when all the sweeps in the DMRG step have been completed, the grouped MPS $M^{[1]}_{t  L_{\mathrm{max}}}, \dots, M^{[g]}_{t  L_{\mathrm{max}}}$ that approximates the state of the circuit after $t  L_{\mathrm{max}}$ layers is decomposed back into a normal MPS via SVD, and employed as the initial state of the circuit in the following iteration. The maximum bond dimension $\chi_{\mathrm{max}} ^{\mathrm{SVD}}$ used in this additional decomposition step is generally chosen much larger than $\chi_{\mathrm{max}} ^{\mathrm{DMRG}}$ to encode as much information as possible.

\section{Numerical results}
\label{sec:results}

\subsection{Tested quantum circuits}
\label{sec:circuits}

To evaluate the performance of the optimized tensor network algorithms, we conduct numerical experiments on two distinct circuit structures. As mentioned in the previous section, our goal is to test the methods on circuits that both generate substantial entanglement and exhibit irregular structures. 

We begin by considering a circuit structure inspired by random circuit sampling~\cite{Aaronson2017QSComplexity,Bouland2019RCS}, where each circuit not only contains random gates but it is also assembled randomly, similar to the design of local random quantum circuits~\cite{Fisher2023RSQC, Harrow2009RSQC2, Brandao2014LocalRQC, mitsuhashi2024unitarydesignssymmetriclocal, mitsuhashi2024characterizationrandomnessquantumcircuits, suzuki2024globalrandomnessrandomlocal}. We refer to this circuit as a ``random-structured quantum circuit''. After choosing the number of qubits and layers, a random-structured quantum circuit is generated by applying either a random single-qubit gate or a random two-qubit gate with a uniform probability $p = 0.5$, until all layers are completed.

\begin{figure*}[ht!]
  \centering
  \includegraphics[width=\textwidth]{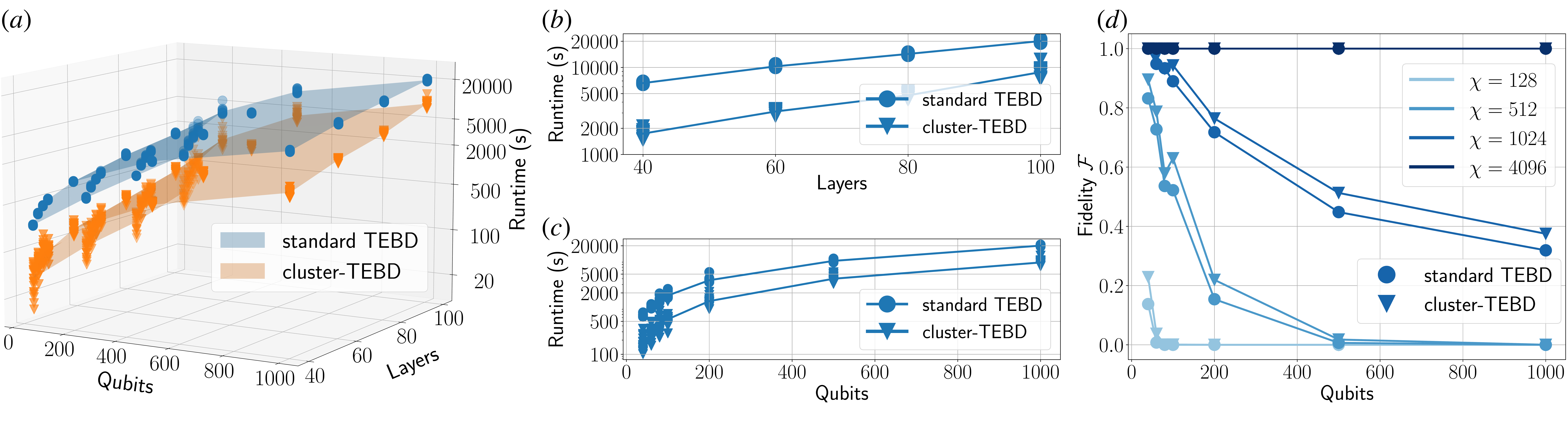}
  \caption{Comparison of cluster-TEBD and TEBD in simulations of Clifford random-structured quantum circuits. \hyperref[fig:benchmarks_cTEBD_clifford]{(a)}~Runtime, in logarithmic scale, of standard TEBD and cluster-TEBD simulations of Clifford random-structured quantum circuits with maximum bond dimension $\chi_{\mathrm{max}}=4096$. \hyperref[fig:benchmarks_cTEBD_clifford]{(b)}~Runtime, in logarithmic scale, of standard TEBD and cluster-TEBD simulations of $1000$-qubit Clifford random-structured quantum circuits with bond dimension $4096$ and varying number of layers. \hyperref[fig:benchmarks_cTEBD_clifford]{(c)}~Runtime, in logarithmic scale, of standard TEBD and cluster-TEBD simulations of $100$-layer Clifford random-structured quantum circuits with bond dimension $4096$ and varying number of qubits. \hyperref[fig:benchmarks_cTEBD_clifford]{(d)}~Fidelity of standard TEBD and cluster-TEBD simulations of $100$-layer Clifford random-structured quantum circuits with varying number of qubits.}
    \label{fig:benchmarks_cTEBD_clifford}
\end{figure*}

The initial state is simply a product of $\ket{0}$s, for all the qubits in the system. We consider two variations of these random-structured quantum circuits: (1) in the first version, single-qubit gates are randomly selected from the set $\left\{H, X, Y, Z\right\}$ and two-qubit gates from the set $\left\{CNOT, C\text{-}Y, C\text{-}Z, SWAP\right\}$. Since all these gates belong to the Clifford group, all the circuits are stabilizer circuits~\cite{Gottesman1998GKTheorem}; (2) in the second version, the sets are extended, adding the non-Clifford gates $\left\{T, P(\frac{3\pi}{4}), \sqrt[4]{X}, \sqrt{W}\right\}$ to the set of single-qubit gates, with $\sqrt{W} = \frac{1}{\sqrt{2}}\left(X + Y\right)$, and the gates $\left\{C\text{-}H, C\text{-}S, C\text{-}T, \sqrt{SWAP}\right\}$ to the set of two-qubit gates. This version introduces a $50\%$ ratio of nonstabilizer elements in the circuits, increasing the complexity of the simulations~\cite{Gottesman1998GKTheorem,Aaronson2004MagicExponential}. For a more detailed description of the gates employed in the simulated circuits, see Appendix~\ref{app:gates}.

In addition to random-structured quantum circuits, we assess the performance of our optimized algorithms on another class of quantum circuits, specifically the implementation of Shor's algorithm for factoring~\cite{Shor1999}, which serves as a representative instance of quantum algorithms. Aside from the classical pre and postprocessing parts of the procedure, the quantum part, which we simulate using tensor network methods, essentially finds the order of $a$ modulo $N$, with $a$ and $N$ positive integers given as input~\cite{NielsenChuang2010}. The central piece of the quantum algorithm is the unitary $U$, which multiplies a positive integer $x$ by $a \hspace{-1mm} \mod N$. Any integer $x$ can be mapped to a quantum state $\ket{x}$ by computing its $n$-bit binary representation $x_0x_1\cdots x_{n-1}$ and assigning each bit to a different qubit: $\ket{x} = \ket{x_0}\otimes\ket{x_1}\otimes\cdots\otimes\ket{x_{n-1}}$. Then, the unitary performs the transformation $U\ket{x} = \ket{ax \!\!\mod N}$. This unitary can be decomposed into a combination of single-, two- and three-qubit gates, following Beauregard's implementation of the controlled modular multiplier~\cite{Beauregard2003ShorCircuit}, based on Draper's quantum addition circuit~\cite{Draper2000QAddition}. The long range interactions, i.e., gates that operate on non-neighboring qubits, are implemented via the application of $SWAP$ gates to draw the qubits close to each other and bring them back to their original position after applying the gate. 

While random-structured quantum circuits are dense, with layers filled with gates, this implementation of Shor's algorithm features a significant number of layers with very few gates per layer. Moreover, gates within similar ranges of layers are localized, meaning they operate on subsets of qubits that are in close proximity, leaving the remaining qubits of the system unaffected. Our approach of clustering together qubits entangled by the evolution proves particularly advantageous in this context of sparse layers, as it is sufficient to focus on a localized section of the state. The broader the range of layers in which these localized operations occur, the greater the number of layers that can be simulated simultaneously within a single cluster. Let us stress that this structure is not a specific feature of Shor's algorithm, but it appears in other known quantum algorithms~\cite{Chen2023QFTSmallEntanglement, Niedermeier2024QalgsTN, Arnault2024typologyquantumalgorithms}.

We run all our benchmarks on the CINECA supercomputer LEONARDO, where we employ one node for each simulated circuit and for each algorithm~\cite{LeonardoCINECA}. In particular, we employ Booster nodes, each with 32 cores and 512GB of RAM, where all resources are exclusively dedicated to each simulation to guarantee reliable measures for the runtime.

\subsection{Cluster-TEBD versus standard TEBD}

In this section, we compare the performances of cluster-TEBD and standard TEBD algorithms in terms of runtime and fidelity for the two aforementioned variations of random-structured quantum circuits, i.e., with Clifford and non-Clifford gates, respectively.

\subsubsection{Clifford random-structured quantum circuits}

We start by showing the results for Clifford random-structured quantum circuits of various sizes. More specifically, we select a number of qubits spanning from $40$ up to $1000$, and circuit depths of $40$, $60$, $80$, and $100$ layers. In the simulations, we vary the maximum bond dimension  $\chi_{\mathrm{max}}$ of the MPS from $128$ to $4096$, while the parameter $Q_{\mathrm{max}}$ is set at $27$. 

As the simulated circuits contain randomness in the distribution of gates, each realization may exhibit different runtime and fidelities of the final states. Consequently, for each circuit size and bond dimension, we repeat the simulation for a total of $20$ samples and average the results. 

Fig.~\ref{fig:benchmarks_cTEBD_clifford} summarizes the results of our benchmarks for Clifford random-structured quantum circuits. The three-dimensional (3D) plot in Fig.~\hyperref[fig:benchmarks_cTEBD_clifford]{8(a)} shows, in logarithmic scale, the runtime of cluster-TEBD and standard TEBD for bond dimension~$4096$, which assures $\mathcal{F} \simeq 1$ for all circuits. The plot clearly shows that the runtime of cluster-TEBD is constantly lower than that of standard TEBD, meaning that, for the considered circuit types and sizes, cluster-TEBD offers a considerable speedup over standard TEBD. If we quantify this speedup by computing the ratio between standard and cluster-TEBD runtimes, we find out that cluster-TEBD is at least $2.276$~times faster than standard TEBD, with a peak speedup of more than $6$~times in the case of $40$~qubit, $40$~layer Clifford random-structured quantum circuits. Figs.~\hyperref[fig:benchmarks_cTEBD_clifford]{8(b)}-\hyperref[fig:benchmarks_cTEBD_clifford]{8(c)} are insets of the 3D plot, with the number of qubits fixed at 1000 and the number of layers fixed at 100, respectively.

We notice that, in general, the speedup does not vary as much with the number of qubits as it does with the circuit depth. This is expected, as chaotic quantum dynamics, such as in random quantum circuits, can generate a large amount of entanglement and lead to rapid entanglement growth. If the bond dimensions increase quickly enough to follow the entanglement growth as much as possible, the clusters that can be formed become smaller, and, consequently, fewer layers can be exactly contracted simultaneously. 

Fig.~\hyperref[fig:benchmarks_cTEBD_clifford]{8(d)} shows the fidelity of the final states obtained from the standard TEBD and cluster-TEBD algorithms when simulating Clifford random-structured quantum circuits with $100$ layers, plotted as a function of the number of qubits in each circuit, for different values of $\chi_{\mathrm{max}}$. It is possible to notice that using $\chi_{\mathrm{max}} = 4096$ as maximum bond dimension guarantees fidelity $\mathcal{F} \simeq 1$ even for the largest circuit analyzed in this work. Notably, using $\chi_{\mathrm{max}} = 128$ produces a fidelity close to $0$ for all Clifford random-structured quantum circuits with $100$ gate layers, while $\chi_{\mathrm{max}} = 512$ and $\chi_{\mathrm{max}} = 1024$ only get slightly close to fidelity $1$ for circuits with tens of qubits. Additionally, while the fidelities for standard TEBD and cluster-TEBD are around the same order of magnitude at every point, we observe an improvement in the fidelity for cluster-TEBD simulations, which may be attributed to the larger number of layers simulated exactly, without any SVD or truncation step.

\subsubsection{Non-Clifford random-structured quantum circuits}

\begin{figure*}[ht!]
  \centering
  \includegraphics[width=\textwidth]{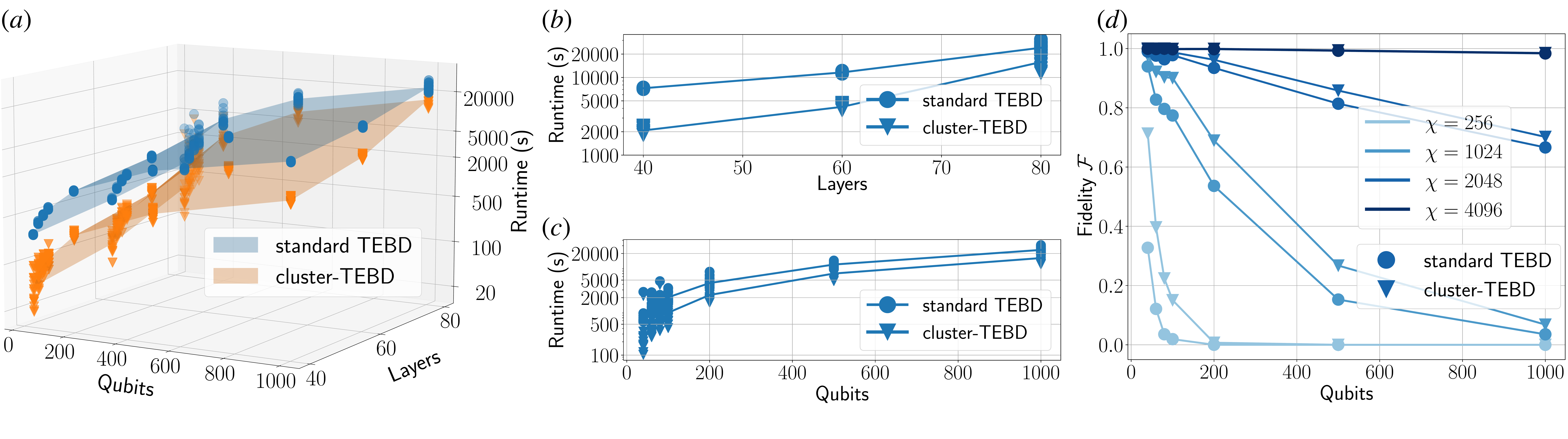}
  \caption{Comparison of cluster-TEBD and TEBD in simulations of non-Clifford random-structured quantum circuits. \hyperref[fig:benchmarks_cTEBD_nonclifford]{(a)}~Runtime, in logarithmic scale, of standard TEBD and cluster-TEBD simulations of non-Clifford random-structured quantum circuits with maximum bond dimension $\chi_{\mathrm{max}}=4096$ \hyperref[fig:benchmarks_cTEBD_nonclifford]{(b)}~Runtime, in logarithmic scale, of standard TEBD and cluster-TEBD simulations of $1000$-qubit non-Clifford random-structured quantum circuits with maximum bond dimension $\chi_{\mathrm{max}}=4096$ and varying number of layers. \hyperref[fig:benchmarks_cTEBD_nonclifford]{(c)}~Runtime, in logarithmic scale, of standard TEBD and cluster-TEBD exact simulations of $80$-layer non-Clifford random-structured quantum circuits with bond dimension $4096$ and varying number of qubits. \hyperref[fig:benchmarks_cTEBD_nonclifford]{(d)}~Fidelity of standard TEBD and cluster-TEBD simulations of $80$-layer non-Clifford random-structured quantum circuits with varying number of qubits.}
  \label{fig:benchmarks_cTEBD_nonclifford}
\end{figure*}

We now benchmark random-structured quantum circuits again, this time with a $50\%$ probability of introducing non-Clifford gates during the evolution, to test the performance of cluster-TEBD on nonstabilizer circuits. We use the same number of qubits and bond dimensions as in the Clifford case, stopping at a circuit depth of $80$ layers. This is because, even with the largest bond dimension used in our simulations, the final fidelity for non-Clifford random-structured quantum circuits with more than $80$ gate layers turns out to be relatively low for both standard and cluster-TEBD. However, a maximum bond dimension $\chi_{\mathrm{max}}=4096$ can still accurately represent the final state of all analyzed circuits of this kind with up to $80$ layers, achieving a fidelity of $\mathcal{F} \simeq 1$.

Fig.~\ref{fig:benchmarks_cTEBD_nonclifford} summarizes the results of our benchmarks, where we average out the runtime and fidelity results of $20$ different samples. Looking at the 3D plot in Fig.~\hyperref[fig:benchmarks_cTEBD_nonclifford]{9(a)}, showing runtimes, in logarithmic scale, for simulations with $\chi_{\mathrm{max}}=4096$, it is evident that the cluster-TEBD runtime is again constantly lower than that of standard TEBD for the considered circuit sizes. We observe that non-Clifford gates increase entanglement, and consequently the bond dimension, more rapidly than Clifford gates. As a result, the number of clustered qubits decreases more quickly compared to stabilizer circuits, reducing the effectiveness of cluster-TEBD slightly. Nevertheless, cluster-TEBD still presents a speedup compared to standard TEBD, going, on average, from $1.5$, for circuits with $1000$ qubits and $80$ layers, up to $5.5$, for circuits with $40$ qubits and $40$ layers.

The increased simulation difficulty of the nonstabilizer version of random-structured quantum circuits is confirmed by the fidelity plot in Fig.~\hyperref[fig:benchmarks_cTEBD_nonclifford]{9(d)}. While representing the final state of these nonstabilizer circuits with intermediate bond dimensions is evidently much harder than in the stabilizer case, $\chi_{\mathrm{max}} = 4096$ again allows us to simulate them exactly. Another interesting feature to note is that the cluster-TEBD advantage in improving final fidelity for low bond dimensions is even more pronounced with non-Clifford gates compared to the Clifford case. This may be due to the larger number of non-Clifford gates being contracted exactly in the cluster-TEBD procedure.

\subsubsection{Shor's algorithm}

\begin{figure}[t]
  \includegraphics[width=8.66cm]{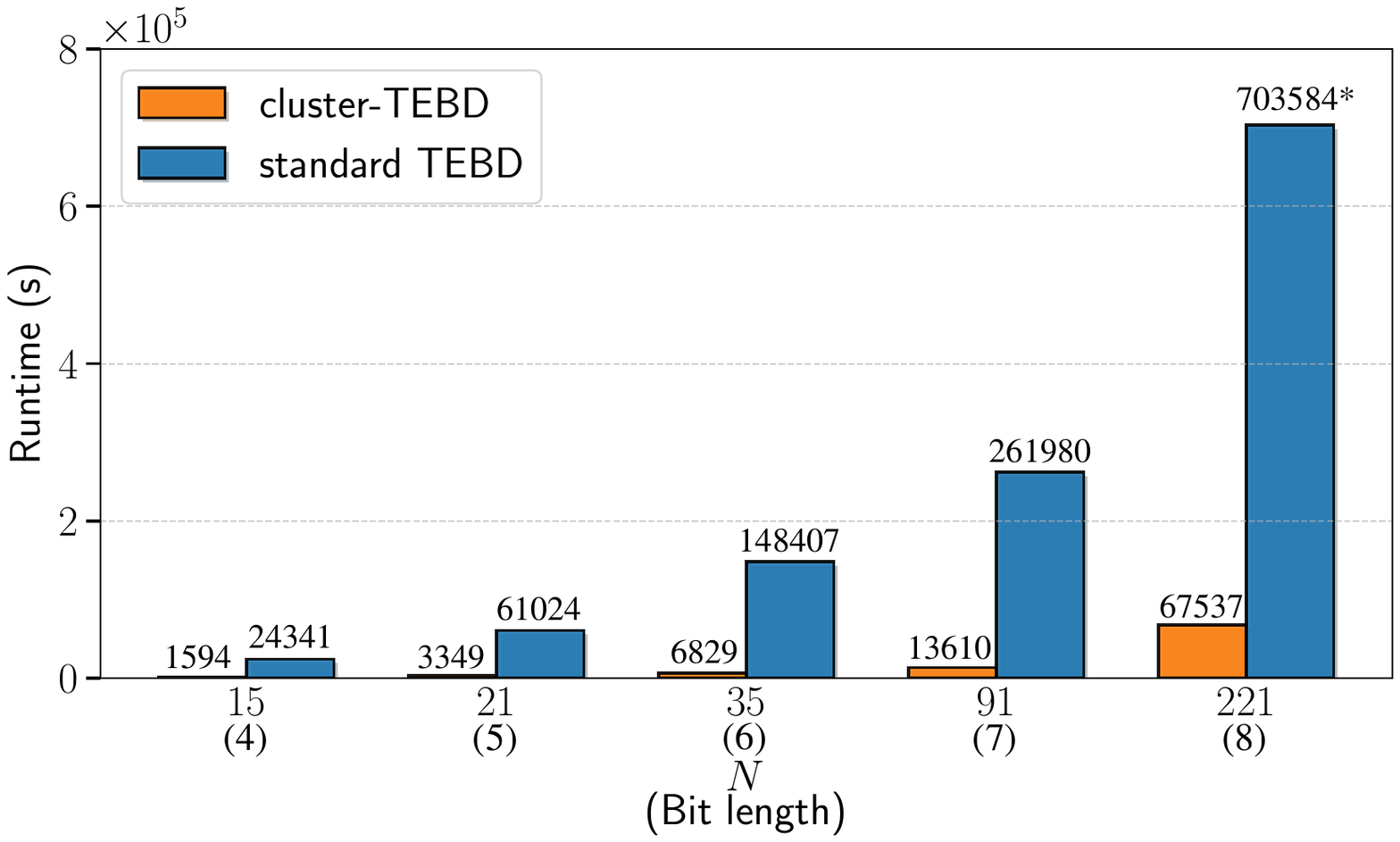}
  \caption{Runtime of standard TEBD and cluster-TEBD simulations of Shor's algorithm for factoring numbers of increasing bit length, from $15$ (four-bit number) to $221$ (eight-bit number). $^*$The standard TEBD runtime provided for $N = 221$ is an extrapolated estimate, as the simulation reached the maximum wall time of the supercomputer; the estimate is based on the observed progress prior to termination.}
  \label{fig:shor}
\end{figure}

The cluster-TEBD algorithm demonstrates significant advantages when applied to the simulation of random-structured quantum circuits, even as the bond dimension quickly becomes very large. In the following, we assess its performance in simulating a representative example of quantum algorithms, where we anticipate that the bond dimension will remain more constrained throughout the evolution~\cite{Chen2023QFTSmallEntanglement, Niedermeier2024QalgsTN}. In this scenario, we expect to group large clusters of entanglement throughout the whole simulation process, leading to improved overall performance.

Shor's algorithm for factoring constitutes a great example of such a task, due to the sparse presence of gates in each layer. Even though the depth of these circuits is of the order of tens, or even hundreds of thousands gate layers, each of these layers contains a limited amount of gates. This is because of the structure of the algorithm itself, which is mainly composed of rotations and $CNOT$ gates on consecutive qubit pairs. Additionally, the presence of long range interactions poses a challenge that can be solved in two possible ways. The first approach is to replace the gate with a multisite MPO, assuming an efficient MPO representation of the gate exists. The second approach -- which we adopt in this work -- involves the application of consecutive $SWAP$ gates, each belonging to a single layer, to operate on distant qubits. 

We simulate Shor's algorithm for factoring the numbers $15, 21, 35, 91, 221$, with adequate precision values. Following Beauregard's implementation of Shor's algorithm~\cite{Beauregard2003ShorCircuit}, the quantum circuits for factoring these numbers display the following characteristics:
\begin{itemize}
    \item For $N = 15$ with $\epsilon = 10^{-3}$: The circuit requires $29$ qubits and $65\,594$ gates, arranged in $45\,420$ layers.
    \item For $N = 21$ with $\epsilon = 10^{-4}$: The circuit requires $37$ qubits and $142\,106$ gates, arranged in $100\,847$ layers.
    \item For $N = 35$ with $\epsilon = 10^{-5}$: The circuit requires $44$ qubits and $257\,841$ gates, arranged in $188\,282$ layers.
    \item For $N = 91$ with $\epsilon = 10^{-5}$: The circuit requires $48$ qubits and $388\,432$ gates, arranged in $292\,635$ layers.
    \item For $N = 221$ with $\epsilon = 10^{-9}$: The circuit requires $65$ qubits and $818\,434$ gates, arranged in $617\,553$ layers.
\end{itemize}
Fig.~\ref{fig:shor} shows the runtime of cluster-TEBD and standard TEBD for simulating this quantum circuit implementation of Shor's algorithm for the specified numbers. The cluster-TEBD runtime is significantly lower than standard TEBD's, as in the case of random-structured quantum circuits, but the speedup in this case is much higher -- cluster-TEBD is between $10$ and $20$ times faster -- and more consistent across the analyzed factorization problems.

This speedup underscores the significant advantage of using cluster-TEBD for this type of quantum circuits, providing substantial improvements in overall performance.

\subsection{DMRG with higher number of layers}
\label{sec:dmrg_res}

Thanks to the adaptive clustering algorithm introduced in Sec.~\ref{subsec:hdmrg}, which generates the grouped MPS according to the circuit structure, it becomes possible to employ the DMRG algorithm to simulate quantum circuits with irregular structures. This allows us to run benchmarks for DMRG on random-structured quantum circuits. In particular, we test the nonstabilizer variation, choosing sizes of $60$, $80$, $100$ qubits, and $60$, $80$ layers, with the goal of understanding how many layers can be exactly contracted in a single step. While in the original work by Ayral \textit{et al.}~\cite{Ayral2023DMRG_QC} the largest number of layers $L_{\mathrm{max}}$ in each iteration employed was $4$, we vary $L_{\mathrm{max}}$ from $2$ to $20$. 

\begin{figure}[t]
      \centering
      \includegraphics[width=8.08cm]{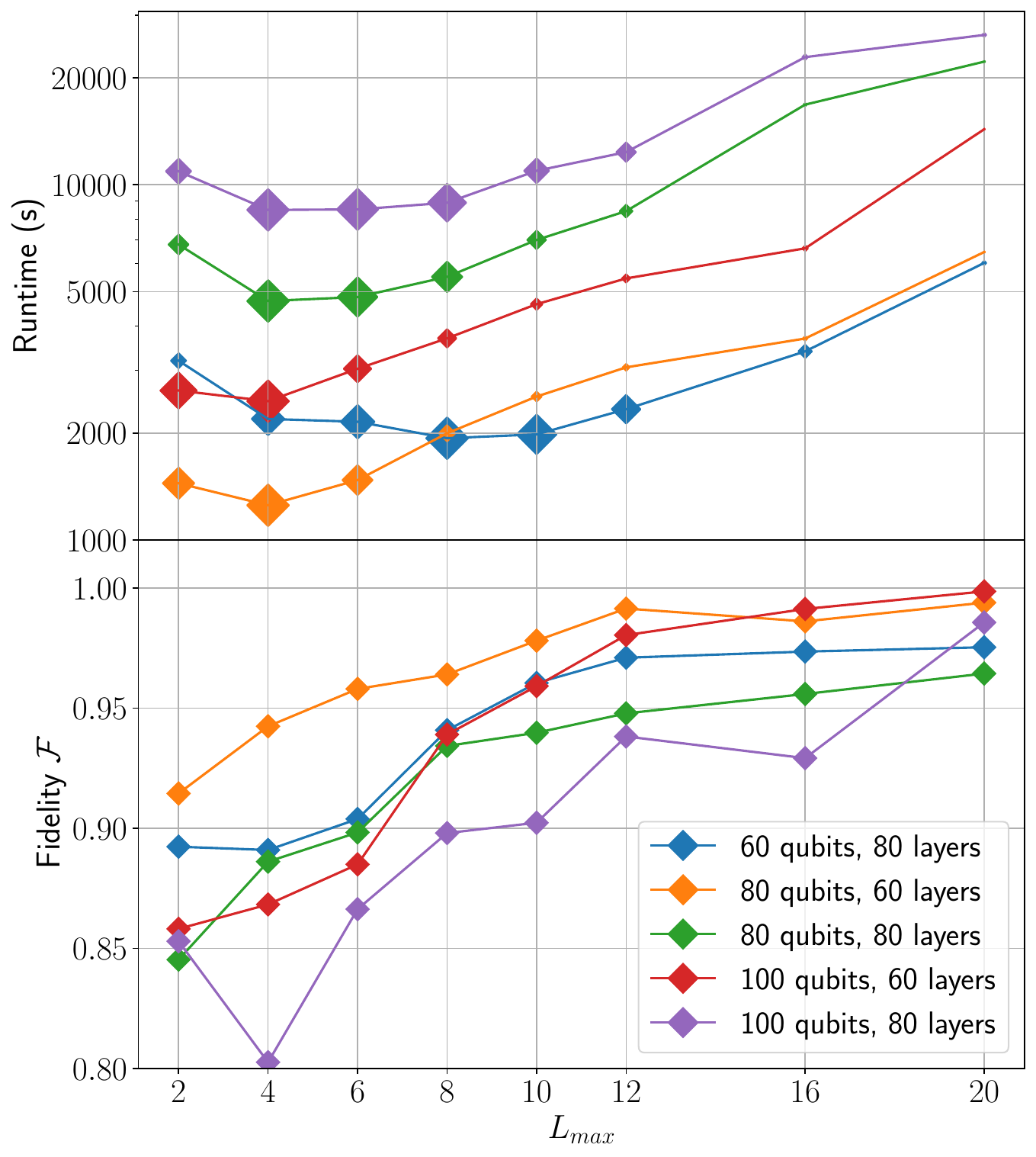}
  \caption{Results of DMRG benchmarks on 20 realizations of non-Clifford random-structured quantum circuits of various sizes with maximum bond dimensions $\chi_{\mathrm{max}} ^{\mathrm{DMRG}} = 256$ and $\chi_{\mathrm{max}} ^{\mathrm{SVD}} = 4096$, and $n_s = 3$ maximum sweeps. Top:~Average runtime of DMRG simulations as a function of the maximum number of layers per step $L_{\mathrm{max}}$. Bottom:~Average fidelity of DMRG simulations as a function of the maximum number of layers per step $L_{\mathrm{max}}$.}
  \label{fig:dmrg_results}
\end{figure}

The top plot in Fig.~\ref{fig:dmrg_results} shows the average runtime of DMRG for 20 non-Clifford circuits of various sizes as a function of $L_{\mathrm{max}}$, fixing the maximum number of sweeps per iteration at $n_s = 3$, the maximum size of a qubit cluster $Q_{\mathrm{max}} = 20$, and the maximum bond dimensions at $\chi_{\mathrm{max}} ^{\mathrm{DMRG}} = 256$ and $\chi_{\mathrm{max}} ^{\mathrm{SVD}} = 4096$. For each circuit, the size of markers identifies how close each point is to the global runtime minimum. The plot shows that, while the value $L_{\mathrm{max}} = 4$ selected in the original work is a very reasonable choice, the actual minimum runtime for varying $L_{\mathrm{max}}$ changes depending on the circuit. On average, the minimum is at 4 for the 80-qubit, 60-layer and the 100-qubit, 60-layer circuits, but it sits at 6 for the 100-qubit, 80-layer and the 80-qubit, 80-layer circuits, and at 10 for the 60-qubit, 80-layer circuit. Moreover, while a minimum runtime can be identified, there are multiple $L_{\mathrm{max}}$ values for each circuit that guarantee a runtime close enough to the minimum. In general, for all circuit sizes, we can identify a pattern of ``wells'' of runtime minima, i.e., a subset of values for $L_{\mathrm{max}}$ where the runtime is close to the minimum. This subset is not necessarily around $L_{\mathrm{max}} = 4$, but it depends on the circuit size, the amount of entanglement in the evolution and the maximum bond dimension $\chi_{\mathrm{max}} ^{\mathrm{DMRG}}$ used throughout the simulation.

In the bottom plot from Fig.~\ref{fig:dmrg_results}, we show the evolution of the average fidelity across the simulated circuit sizes as a function of $L_{\mathrm{max}}$, demonstrating an advantage for larger $L_{\mathrm{max}}$ values. This improvement in fidelity is due to several reasons. First, the dynamical adaptive grouping routine requires the grouped MPS to be converted back into a standard MPS with successive SVDs. Setting a lower $L_{\mathrm{max}}$ value increases the frequency of these decompositions throughout the algorithm. Since each decomposition step can introduce fidelity loss through truncation, a higher $L_{\mathrm{max}}$ value helps to reduce these potential reductions. Additionally, for growing $L_{\mathrm{max}}$, the number of gate layers contracted exactly within a single iteration rises accordingly.

Appendix~\ref{app:dmrg} provides additional results for the DMRG algorithm, where we compare the runtime performance of using the dynamical adaptive grouping routine introduced in this work against a fixed grouping.

\section{Discussion}
\label{sec:discussion}
In this work, we have optimized tensor network algorithms for simulating quantum circuits with finite fidelity by implementing entanglement clustering protocols. We designed a variation of the TEBD algorithm, called the ``cluster-TEBD'' algorithm, which allows the exact contraction of multiple layers of quantum gates in a single step. This algorithm employs an adaptive clustering technique that groups gates in different layers based on the arrangement of entangling operations, increasing each cluster's size until the selected layers can be contracted exactly. Moreover, we implemented an automatic clustering scheme for the DMRG algorithm, using hypergraph partitioning~\cite{Schlag2022KaHyPar}, where the qubit grouping adaptively changes in each iteration according to the arrangement of entangling gates in the simulated circuit portion. These enhancements allowed efficient simulations of quantum circuits where entangling operations are scattered across a large chunk of gate layers, rather than condensed into few layers.

Our results demonstrate that cluster-TEBD significantly outperforms standard TEBD in both runtime and fidelity when simulating quantum circuits with random gates and structure. Furthermore, we have shown that this advantage is not only maintained but also considerably enhanced when applied to the simulation of a more practical task, specifically Shor's quantum algorithm for factoring. Additionally, using the adaptive clustering algorithm in DMRG, we have focused on studying the effect of the choice of the maximum number of contracted layers per iteration $L_{\mathrm{max}}$. We have demonstrated that, for a given quantum circuit to simulate, there exists a range of valid values for $L_{\mathrm{max}}$, even larger than what was considered in previous works~\cite{Ayral2023DMRG_QC}. Moreover, we have verified that, using the dynamical adaptive grouping routine, the final fidelity of the simulation increases with $L_{\mathrm{max}}$.

The framework introduced in this work could be adapted to other tensor network ansätze beyond MPS. For instance, it could be extended to tree tensor networks~\cite{Shi2006TTN, 10.21468/SciPostPhysLectNotes.8} and hybrid quantum-classical structures~\cite{PhysRevLett.127.040501, schuhmacher2024hybridtreetensornetworks}, combining them with adaptive techniques to build the network according to the structure of the evolution~\cite{Ferrari2022AdaptiveTTN, Seitz2023QC_TTN}. Additionally, our dynamical clustering methods can be used in combination with belief propagation gauging, which has recently proven extremely effective in tensor network simulations with Projected Entangled Pair States~\cite{Tindall2023BPGauging, Tindall2025_2DTNs, Rudolph2025_2DQC}. These modifications would allow to extend our numerical methods to simulate quantum circuits in 2D qubit geometries. 

Another promising aspect is the potential for parallelization in cluster-TEBD, as the contraction of each cluster can be performed independently before orthogonalizing the state to measure observables. Introducing large-scale parallelism, such as MPI-based methods, could further enhance the performance~\cite{gabriel04:_open_mpi}. 

Cluster-TEBD and enhanced DMRG algorithms could be used for solving complex physical tasks in the near future, such as studying the behavior of NISQ quantum computers, supporting their developments and strengthening the crucial synergy between tensor networks and quantum computing.

\section*{Acknowledgments}
The authors acknowledge financial support from the following institutions: the European Union (EU), via the NextGenerationEU projects ``CN00000013 - Italian Research Center on HPC, Big Data and Quantum Computing (ICSC)'' and ``PE0000023  - National Quantum Science and Technology Institute (NQSTI)''; the European Union's Horizon Europe research and innovation programme (Quantum Flagship) under the projects PASQuanS2 (Grant Agreement No.~101113690) and EuRyQa (Grant Agreement No.~101070144); the Italian Ministry of University and Research (MUR), via PRIN2022 project ``2022NZP4T3 - QUEXO'' (CUP D53D23002850006), and via the Department of Excellence grants 2023-2027 project ``Quantum Sensing and Modelling for One-Health (QuaSiModO)''; the Italian Istituto Nazionale di Fisica Nucleare (INFN), via the projects QUANTUM and NPQCD; the University of Bari, via the 2023-UNBACLE-0244025 grant. This research is part of the Munich Quantum Valley, which was supported by the Bavarian state government with funds from the Hightech Agenda Bayern Plus. We acknowledge computational resources by CINECA and the University of Bari and INFN cluster RECAS~\cite{ReCaS}. The numerical simulations were performed using the following libraries: \texttt{ITensors.jl}~\cite{Fishman2022ITensors}, for management of tensors and MPS, \texttt{opt\_einsum}~\cite{Smith2018opt_einsum}, and \texttt{cotengra}~\cite{Gray2021HyperExact}, for heuristics and optimized exact contractions.

\section*{Data availability}
The data that support the findings of this article are not publicly available. The data are available from
the authors upon reasonable request.

\appendix

\section{Gates used in random-structured quantum circuits}
\label{app:gates}
In this Appendix, we define the gates used in random-structured quantum circuits. Let us start with the Clifford single-qubit gates, which are simply
\begin{equation}
  H := \frac{1}{\sqrt{2}}\begin{pmatrix}
    1 & 1 \\ 1 & -1
\end{pmatrix}, \hspace{2mm} X := \begin{pmatrix}
    0 & 1 \\ 1 & 0
\end{pmatrix} ,   
\end{equation}
\begin{equation}
 Y := \begin{pmatrix}
    0 & -i \\ i & 0
\end{pmatrix}, \hspace{2mm} Z := \begin{pmatrix}
    1 & 0 \\ 0 & -1
\end{pmatrix} .    
\end{equation}
The Clifford two-qubit gates are
\begin{equation}
CNOT := \begin{pmatrix}
    1 & 0 & 0 & 0 \\ 0 & 1 & 0 & 0 \\ 0 & 0 & 0 & 1 \\ 0 & 0 & 1 & 0
\end{pmatrix}, \hspace{2mm} C\text{-}Y := \begin{pmatrix}
    1 & 0 & 0 & 0 \\ 0 & 1 & 0 & 0 \\ 0 & 0 & 0 & -i \\ 0 & 0 & i & 0
\end{pmatrix},    
\end{equation}
\begin{equation}
 C\text{-}Z := \begin{pmatrix}
    1 & 0 & 0 & 0 \\ 0 & 1 & 0 & 0 \\ 0 & 0 & 1 & 0 \\ 0 & 0 & 0 & -1
\end{pmatrix}, \hspace{2mm} SWAP := \begin{pmatrix}
    1 & 0 & 0 & 0 \\ 0 & 0 & 1 & 0 \\ 0 & 1 & 0 & 0 \\ 0 & 0 & 0 & 1
\end{pmatrix}.   
\end{equation}

Let us now move on to non-Clifford gates. Let us first define the phase gate $P(\varphi)$:
\begin{equation}
  P(\varphi) := \begin{pmatrix}
    1 & 0 \\ 0 & e^{i\varphi}
\end{pmatrix}.  
\end{equation}
The non-Clifford single-qubit gates we used are
\begin{equation}
  T := P(\frac{\pi}{4}) = \begin{pmatrix}
    1 & 0 \\ 0 & e^{i\frac{\pi}{4}}
\end{pmatrix}, \hspace{2mm} P\left(\frac{3\pi}{4}\right) = \begin{pmatrix}
    1 & 0 \\ 0 & e^{i\frac{3\pi}{4}}
\end{pmatrix},  
\end{equation}
\begin{equation}
   \sqrt[4]{X} = e^{i\frac{\pi}{8}}\begin{pmatrix}
    \cos\left(\frac{\pi}{8}\right) & -i\sin\left(\frac{\pi}{8}\right) \\ -i\sin\left(\frac{\pi}{8}\right) & \cos\left(\frac{\pi}{8}\right)
\end{pmatrix}, 
\end{equation}
\begin{equation}
 \sqrt{W} = \sqrt{\frac{1}{\sqrt{2}}\left(X + Y\right)} = \frac{1}{\sqrt{2}}\begin{pmatrix}
    \frac{1}{e^{i\frac{\pi}{4}}} & -i \\ 1 & \frac{1}{e^{i\frac{\pi}{4}}}
\end{pmatrix}.   
\end{equation}
Finally, the employed non-Clifford two-qubit gates are
\begin{equation}
C\text{-}H = \begin{pmatrix}
    1 & 0 & 0 & 0 \\ 0 & 1 & 0 & 0 \\ 0 & 0 & \frac{1}{\sqrt{2}} & \frac{1}{\sqrt{2}} \\ 0 & 0 & \frac{1}{\sqrt{2}} & -\frac{1}{\sqrt{2}}
\end{pmatrix},    
\end{equation}
\begin{equation}
  C\text{-}S := C\text{-}P(\frac{\pi}{2}) \begin{pmatrix}
    1 & 0 & 0 & 0 \\ 0 & 1 & 0 & 0 \\ 0 & 0 & 1 & 0 \\ 0 & 0 & 0 & i
\end{pmatrix},  
\end{equation}
\begin{equation}
   C\text{-}T = \begin{pmatrix}
    1 & 0 & 0 & 0 \\ 0 & 1 & 0 & 0 \\ 0 & 0 & 1 & 0 \\ 0 & 0 & 0 & e^{i\frac{\pi}{4}}
\end{pmatrix}, 
\end{equation}
\begin{equation}
 \sqrt{SWAP} = \begin{pmatrix}
    1 & 0 & 0 & 0 \\ 0 & \frac{1}{2}(1+i) & \frac{1}{2}(1-i) & 0 \\ 0 & \frac{1}{2}(1-i) & \frac{1}{2}(1+i) & 0 \\ 0 & 0 & 0 & 1
\end{pmatrix}.   
\end{equation}

\section{Further results for DMRG: fixed versus adaptive grouping}
\label{app:dmrg}

\begin{figure}
  \includegraphics[width=8.66cm]{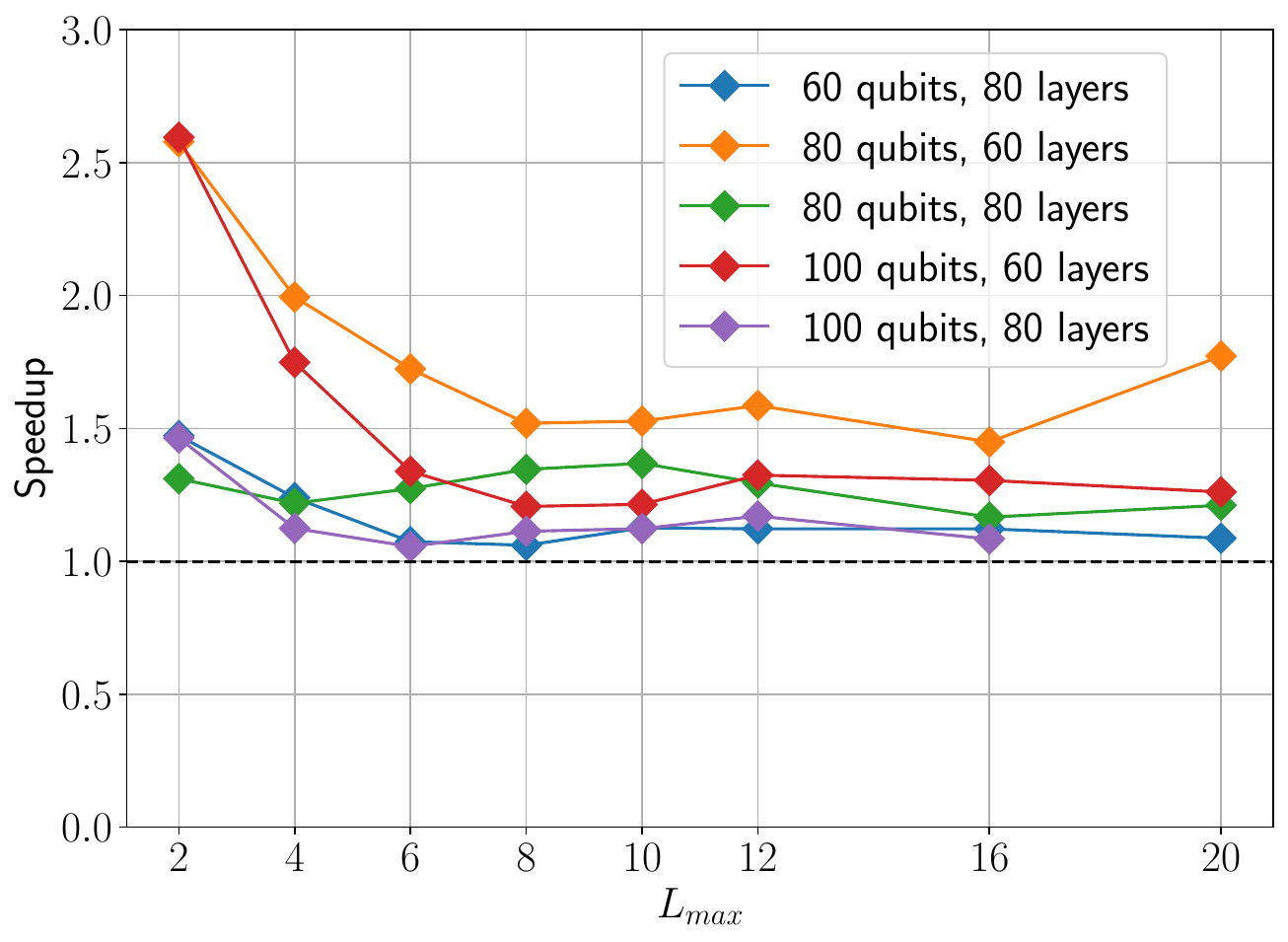}
  \caption{Average speedup of the DMRG algorithm with adaptive grouping against fixed grouping for 20 realizations of non-Clifford random-structured quantum circuits of various sizes.}
  \label{fig:dmrg_speedup}
\end{figure}

In this Appendix, we present additional results from the DMRG benchmarks illustrated in Sec.~\ref{sec:dmrg_res}. In particular, our objective is to test our dynamical adaptive grouping routine by comparing its runtime with a fixed grouping approach. In the fixed grouping implementation, we relax the constraints of the algorithm by determining the grouping based on the gate distribution across the entire circuit prior to the first iteration. This predetermined grouping is then employed across all subsequent DMRG iterations. A key difference with this approach is that, as the grouping is fixed throughout the algorithm, no decomposition steps are required at the beginning of each new iteration.

In Fig.~\ref{fig:dmrg_speedup}, we illustrate the speedup of the DMRG algorithm with adaptive grouping against fixed grouping for 20 different samples of non-Clifford random-structured quantum circuits, with circuit sizes ranging from 60 to 100 qubits and 60 to 80 gate layers. The speedup is computed as follows:
\begin{equation}
 \text{speedup}(Q, L) = \frac{\text{runtime}_{\text{fixed}}(Q, L)}{\text{runtime}_{\text{adaptive}}(Q, L)} \, ,   
\end{equation}
where $\text{runtime}_{\text{fixed}}(Q, L)$ is the runtime for DMRG with fixed grouping for a circuit of $Q$ qubits and $L$ layers, and $\text{runtime}_{\text{adaptive}}(Q, L)$ is the runtime with adaptive grouping. As shown in the figure, all points lie above $1.0$, meaning that using the dynamical adaptive grouping routine in the DMRG algorithm yields a lower runtime on average with respect to fixed grouping. Moreover, the speedup tends to increase for both large and small values of $L_{\mathrm{max}}$, whereas for intermediate values of $L_{\mathrm{max}}$ the speedup is very slight.

\bibliography{bibliography}

\end{document}